\documentclass[a4paper,journal,romanappendices]{IEEEtran}
\usepackage{amssymb}
\usepackage{amsmath}
\usepackage{mathrsfs}
\usepackage{rotating}
\usepackage{overpic}
\usepackage{algorithm}
\usepackage{algorithmic}
\usepackage{multirow}
\usepackage{array}
\usepackage{graphicx}
\usepackage[caption=false]{subfig}
\usepackage{bm}
\usepackage{color}
\usepackage{tikz}
\usepackage{epstopdf}
\usepackage{stfloats}
\usepackage{cite}
\usepackage{tikz}
\usepackage{amsmath}
\usepackage{psfrag}
\usepackage{acronym}
\usepackage{enumerate}
\usetikzlibrary{arrows}
\usetikzlibrary{decorations}

\definecolor{BLUE}{rgb}{0,0,1}

\newtheorem{proposition}{Proposition}
\newtheorem{remark}{Remark}

\newtheorem{definition}{Definition}

\newcommand{\tr}[1]{{\rm Tr}\left\{#1\right\}}

\acrodef{qem}[QEM]{quantum error mitigation}
\acrodef{qpr}[QPR]{quasi-probability representation}
\acrodef{kkt}[KKT]{Karush-Kuhn-Tucker}
\acrodef{nisq}[NISQ]{noisy intermediate-scale quantum}
\acrodef{fir}[FIR]{finite impulse response}
\acrodef{qaoa}[QAOA]{quantum approximate optimization algorithm}
\acrodef{qsa}[QSA]{quantum search algorithm}
\acrodef{qft}[QFT]{quantum Fourier transform}
\acrodef{sts}[STS]{spatio-temporal stabilizer}

\acrodefplural{qsa}[QSA]{quantum search algorithms}
\acrodefplural{sts}[STSs]{spatio-temporal stabilizers}

\usepackage{winsnotation}
\usepackage[braket]{qcircuit}
\usepackage{mathrsfs}
\usepackage{rotating}
\newcommand{\Sop}[1]{{\mathcal{#1}}}

\newtheorem{sufcondition}{Sufficient Condition}

\definecolor{myblue}{rgb}{0.2,0.2,0.7}

\newcommand{\tabincell}[2]{\begin{tabular}{@{}#1@{}}#2\end{tabular}}

\begin{document}
\title{Circuit Symmetry Verification Mitigates Quantum-Domain Impairments}
\author{Yifeng Xiong, Daryus Chandra, \IEEEmembership{Member, IEEE}, Soon Xin Ng, \IEEEmembership{Senior Member, IEEE}, \\ and Lajos Hanzo, \IEEEmembership{Fellow, IEEE}
\thanks{The authors are with the School of Electronics and Computer Science, University of Southampton, SO17 1BJ, Southampton, (UK).}
\thanks{L. Hanzo would like to acknowledge the financial support of the Engineering and Physical Sciences Research Council projects EP/N004558/1, EP/P034284/1, EP/P034284/1, EP/P003990/1 (COALESCE), of the Royal Society's Global Challenges Research Fund Grant as well as of the European Research Council's Advanced Fellow Grant QuantCom. This work is also supported in part by China Scholarship Council (CSC).}
}

\maketitle

\begin{abstract}
State-of-the-art noisy intermediate-scale quantum computers require low-complexity techniques for the mitigation of computational errors inflicted by quantum decoherence. Symmetry verification constitutes a class of quantum error mitigation (QEM) techniques, which distinguishes erroneous computational results from the correct ones by exploiting the intrinsic symmetry of the computational tasks themselves. Inspired by the benefits of quantum switch in the quantum communication theory, we propose beneficial techniques for circuit-oriented symmetry verification that are capable of verifying the commutativity of quantum circuits without the knowledge of the quantum state. In particular, we propose the spatio-temporal stabilizer (STS) technique, which generalizes the  {conventional quantum-domain}  stabilizer formalism to circuit-oriented stabilizers. The applicability and implementational strategies of the proposed techniques are demonstrated by using practical quantum algorithms, including the quantum Fourier transform (QFT) and the quantum approximate optimization algorithm (QAOA).
\end{abstract}


\section{Introduction}
Noisy intermediate-scale quantum computers, exemplified by Google's Sycamore \cite{sycamore} and USTC's Zuchongzhi \cite{zuchongzhi}, are potentially capable of outperforming classical supercomputers on certain specific computational tasks. However, it is widely believed that ubiquitous quantum advantage will only become possible when fault-tolerance \cite{fault_tolerance,qecc,transversal} is achieved, which may not be feasible for noisy intermediate-scale quantum computers due to their limited number of qubits and  {relatively high}  gate error rates. Variational quantum algorithms \cite{vqe,vqe_theory,vqe2,vqe_spectra,vqlinear,sgd_vqa} are thus designed to share their computational tasks between a classical device and a quantum processor, which has the potential of supporting certain practical applications such as molecular simulations and combinatorial optimization \cite{qaoa,performance_qaoa,scalable_simulation,subspace_expansion1}.

One of the important enabling techniques for variational quantum algorithms to become practical is \ac{qem} \cite{qem}, which refers to a class of low-complexity error mitigation techniques that require less qubits than quantum error-correction codes \cite{qecc,qtc1,qtc2,qtecc,qecc_survey}, hence they are particularly suitable for noisy intermediate-scale devices. Existing \ac{qem} methods roughly fall into four categories, namely those based on zero-noise extrapolation \cite{qem,practical_qem,zne1,zne2,zne3}, channel inversion \cite{qem,practical_qem,qem_exp,sof_analysis,ryuji_cost_qem}, machine learning \cite{clifford_regression}, and symmetry verification \cite{sv,sv2,sv3,sv4}, respectively.

Specifically, zero-noise extrapolation methods aim for estimating the true computational result with the aid of several noisy results obtained under different noise levels. Channel inversion methods mitigate the errors by emulating the inverse channels implemented using samples from ``quasi-probability distributions'', which require \textit{a priori} knowledge about the specific channels modelling the impairments of the quantum gates \cite{qem_exp}. Machine learning methods first train statistical models on relatively simple quantum circuits that can be efficiently simulated on classical devices (e.g. Clifford gates), and apply the resultant trained models for mitigating the errors encountered in more sophisticated circuits \cite{clifford_regression}. Symmetry verification methods exploit the symmetries (redundancy) in the computational tasks themselves, and distinguish erroneous results from the correct ones by testing whether the natural symmetries are violated \cite{sv}. The symmetries are typically modelled using the stabilizer formalism. However, they are embedded into the computational tasks themselves rather than those manually designed in quantum error-correction codes. Typically, the number of such intrinsic symmetries is insufficient for identifying and correcting the specific error pattern, hence the computation is often discarded upon detecting a violation of symmetry. In practice, these methods are not necessarily applied in isolation; rather, beneficial combinations have been considered \cite{balint_vd}.

Recently, a new symmetry-aided \ac{qem} method was proposed, known as ``virtual distillation'' \cite{balint_vd,vd2}. This method prepares multiple copies of the quantum circuit to be protected, and verifies the permutation symmetry  {across}  different copies. Exponential accuracy improvement has been observed as the number of copies increases \cite{balint_vd,vd2}. Compared to other existing symmetry verification methods, virtual distillation is more flexible, since the permutation symmetry can be designed by appropriately choosing the number of copies.

From a broader perspective, the virtual distillation method may be viewed as exploiting the spatial consistency among different circuit copies. A natural question that arises is whether we could generalize the idea to the time domain, in the sense that some temporal consistency of the circuit may be verified. This requires a generalization of the conventional state-oriented symmetry verification to \textit{circuit-oriented symmetry verification}.

A related topic, namely the superposition of causal orders \cite{qs_early,qs_comp,qs_prl}, which can be physically realized using the quantum switch of \cite{qs_ieee}, has been investigated from the perspective of quantum communication. Specifically, it has been shown that the capacity of two quantum channels $\Sop{A}$ and $\Sop{B}$ may be improved by producing a coherent superposition between their compositions of different orders, i.e. $\Sop{A}\circ \Sop{B}$ and $\Sop{B}\circ\Sop{A}$ \cite{qs_comm1,qs_comm2,qs_comm3}. More surprisingly, non-zero capacity is achievable even if both the capacity of $\Sop{A}$ and that of $\Sop{B}$ are zero \cite{qs_comm4}. The implementation of the quantum switch relies on a control qubit, the state of which may be used to indicate the commutativity between the composite channels.

In this treatise, we argue that the quantum switch based method can be beneficially used for \ac{qem}, with some modifications. In particular, the quantum switch and its derivations are capable of verifying circuit symmetries such as the commutativity between quantum gates. This is in stark contrast to existing symmetry verification methods relying on stabilizer checks, which aim for verifying the specific properties of quantum states instead of circuits. Against this background, our main contributions are summarized as follows.
\begin{itemize}
\item For quantum circuits consisting of mutually commuting gates, we propose to use the original form of the quantum switch to verify the gate commutativity.
\item For quantum circuits commuting with known operators, especially Pauli operators, we propose a modified quantum switch based method termed as the \ac{sts}, which may be used for detecting and mitigating errors violating the commutativity condition. In contrast to conventional stabilizer-based symmetry verification, \acp{sts} do not depend on the specific quantum state, hence they are more generally applicable.
\item We discuss the practical issues when implementing the \ac{sts} method, including the simultaneous observability of \acp{sts} and their accuracy vs. overhead trade-off. We also provide quantum circuit designs that strike flexible accuracy vs. overhead trade-offs.
\item We demonstrate the usefulness of the \ac{sts} method by applying it to practical quantum algorithms, including the \ac{qft} and the \ac{qaoa}, where the conventional stabilizer checks are not applicable.
\end{itemize}

We organize the rest of this treatise as follows. In Section \ref{sec:symmetries}, we elaborate on the difference between state symmetries and circuit symmetries. Then, in Section \ref{sec:quantum_switch}, we present the implementations of the quantum switch for verifying gate commutativity. For circuits having explicitly known symmetries, we propose the spatio-temporal stabilizers method in Section \ref{sec:sts}. In particular, we present the analysis and the implementation of spatio-temporal stabilizers in Section \ref{ssec:qs_mod} and \ref{ssec:sts}, respectively, followed by our discussions of the associated practical issues, including the simultaneous observability and the accuracy vs. overhead trade-off in Section \ref{ssec:simutaneous_observability} and \ref{ssec:accuracy_vs_overhead}. We then discuss the strategies of applying the method of spatio-temporal stabilizers to practical quantum algorithms in Section \ref{sec:case_study}. Our numerical results are discussed in Section \ref{sec:numerical}, and finally, we conclude in Section \ref{sec:conclusions}.

\section{A General Perspective: Symmetry-Aided \ac{qem} and Signal Processing}
In this section, we introduce the basics of symmetry-aided \ac{qem}, focussing on its deep connections with the concepts in the classical theory of signal processing.

\subsection{Symmetries, Stabilizers, and Subspace Projections}
One of the most widely applied denoising technique in classical signal processing is projecting the observation onto a subspace known as the signal subspace. For example, let us consider the model
\begin{equation}
\M{Y} = \M{H}\M{X}+\M{N},
\end{equation}
where $\M{Y}\in\mathbb{C}^{M\times T}$ denotes the observation containing $T$ independent samples, $\M{H}\in\mathbb{C}^{M\times N}$ represents the (probably known) channel, $\M{X}\in\mathbb{C}^{N\times T}$ is the transmitted signal (typically assumed to have zero mean), and $\M{N}\in\mathbb{C}^{M\times T}$ denotes the noise, whose columns are typically modelled as zero-mean Gaussian random vectors following the distribution $\mathcal{N}(\V{0},\sigma^2\M{I})$. In order to estimate certain parameters related to the channel $\M{H}$ (e.g. direction of arrival), one may first construct the sample covariance matrix
\begin{equation}
\M{R}_{\M{Y}}=\frac{1}{T}\M{Y}\M{Y}^{\rm H},
\end{equation}
which satisfies
\begin{equation}\label{asymptotic_covariance}
\mathbb{E}\{\M{R}_{\M{Y}}\} = \frac{1}{T}\M{H}\mathbb{E}\{\M{X}\M{X}^{\rm H}\}\M{H}^{\rm H}+\sigma^2\M{I}.
\end{equation}

Let us assume that ${\rm rank}(\M{H}\mathbb{E}\{\M{X}\M{X}^{\rm H}\}\M{H}^{\rm H})=r< M$. According to \eqref{asymptotic_covariance}, when the number of samples $T$ is large, the sample covariance matrix $\M{R}_{\M{Y}}$ would contain $(M-r)$ eigenvalues that are numerically close to $\sigma^2$, which correspond to the noise subspace \cite[Section~4.5]{spectral_analysis}. By contrast, other eigenvalues would be larger than $\sigma^2$, and correspond to the specific signal subspace the matrix $\M{H}\mathbb{E}\{\M{X}\M{X}^{\rm H}\}\M{H}^{\rm H}$ resides in. By projecting the sample covariance matrix onto the signal subspace, the deleterious effects of the noise on the estimation performance can be significantly mitigated \cite[Section~4.5]{spectral_analysis}. Under the circumstances where we know in advance certain characteristics of the signal subspace, we may directly apply the corresponding projection operators without resorting to the eigendecomposition of the sample covariance matrix. For example, when the signal is known to be slowly varying, we may apply low-pass filters to mitigate the high-frequency noise.

\begin{figure}
\centering
\includegraphics[width=.26\textwidth]{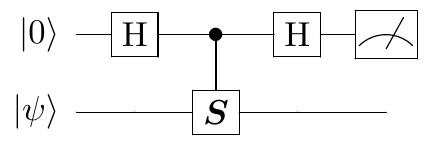}
\caption{A quantum circuit measuring the stabilizer $\M{S}$ of the state $\ket{\psi}$.}
\label{fig:stabilizer_measure}
\end{figure}

Similar principles can also be applied to the quantum domain. In particular, symmetry is a property shared by many practical quantum systems, which may be exploited to identify and mitigate certain sources of error. Roughly speaking, a symmetry of a quantum state refers to a certain transformation under which the state is invariant. More precisely, for a given quantum state $\ket{\psi}$, each type of symmetry is characterized by its associated stabilizer $\M{S}$, satisfying
\begin{equation}\label{stabilizer_def}
\M{S}\ket{\psi}=\ket{\psi},
\end{equation}
implying that $\ket{\psi}$ resides in the invariant subspace corresponding to the eigenvalue $1$ of the stabilizer $\M{S}$. If we have the prior knowledge that a quantum state $\ket{\psi}$ has certain symmetries, we may project the state onto the intersection of the invariant subspaces of all stabilizers, and thus mitigate any error that violates the symmetry conditions. A widely used approach of implementing this projection is to measure the stabilizers (as observables) with the aid of some ancillary qubits (ancillas). As portrayed in Fig.~\ref{fig:stabilizer_measure}, we may project the state $\ket{\psi}$ onto the subspace in which \eqref{stabilizer_def} is satisfied, by discarding the computation upon measuring $\ket{1}$ at the ancilla. \ac{qem} methods exploiting such stabilizer measurements are known as symmetry verification \cite{sv2}.

A more straightforward and more interesting connection between classical subspace projection and quantum symmetry verification is the one exemplified by the recently proposed \ac{qem} method of virtual distillation \cite{balint_vd,vd2}. Specifically, virtual distillation exploits the translation invariance between different copies of the same quantum state. Verifying a slightly modified version of this specific symmetry is effectively equivalent to applying a high-pass filter to the quantum state in its spectral domain, in the sense that the components corresponding to smaller eigenvalues would be further attenuated. We will not present the technical details of this filtering interpretation of virtual distillation in this treatise. Instead, interested readers are referred to \cite{perm_filter}.

\subsection{State Symmetry, Circuit Symmetry, and Differential Modulation}\label{sec:symmetries}
The symmetry discussed in the previous subsection is a property of a quantum state (resp. a signal in the classical domain). However, the information required for further processing is sometimes not represented by the state itself, but it is encoded in the variation between adjacent states (in time) instead.

In the classical domain, a typical example is differential encoding in modulation \cite{dm}. A specific variant of differential modulation, termed as differential space-time modulation \cite{dstm1,dstm2}, generates symbols taking the following form
\begin{equation}
\M{A}_n = \M{U}_n\M{A}_{n-1},
\end{equation}
where $\M{A}_n$ is the symbol transmitted in the $n$-th time slot, but it is the unitary matrix $\M{U}_n$ that actually carries the information. In order to protect the information from the deleterious effect of additive noise, one may encode $\M{U}_n$ using error correction codes.

In the quantum domain, an analogous scenario is the state evolution under the actions on consecutive gates. Upon denoting the quantum state at the $n$-th time instance as $\ket{\psi}_n$, we have
\begin{equation}
\ket{\psi}_n = \M{G}_n\ket{\psi}_{n-1},
\end{equation}
where $\M{G}_n$ is the gate applied at the $n$-th time instance, which also has the mathematical representation of a unitary matrix. Similar to the classical example of differential modulation, we may apply quantum error correction codes to protect the gates $\M{G}_n$. In general, this is equivalent to encode the state $\ket{\psi}_n$ using the same quantum error correction code \cite[Sec.~10.6.2]{ncbook}. However, the situation becomes different when we take into account the intrinsic symmetries of the quantum circuit (i.e. the set of all gates $\M{G}_n$) originated from the computational task itself.

To elaborate, let us consider the simple quantum circuit portrayed in Fig. \ref{fig:symmetric_circuit}. In this diagram, $\Sop{R}_{\rm x}(\cdot)$ denotes a single-qubit X-rotation gate, while $\Sop{R}_{\rm xx}(\cdot)$ denotes a two-qubit XX-rotation gate, which may be mathematically represented as \cite{ncbook}
\begin{equation}
\begin{aligned}
\Sop{R}_{\rm x}(\theta)\ket{\psi}&=\exp\left(-\frac{\imath\theta}{2}\M{X}\right)\ket{\psi},\\
\Sop{R}_{\rm xx}(\theta)\ket{\psi}&=\exp\left(-\frac{\imath\theta}{2}\M{X}\otimes \M{X}\right)\ket{\psi},
\end{aligned}
\end{equation}
where $\M{X}$ denotes the Pauli-X matrix given by
$$
\M{X}=\left[
        \begin{array}{cc}
          0 & 1 \\
          1 & 0 \\
        \end{array}
      \right].
$$

Note that this circuit may be represented by an operator that is diagonal under the X-basis. To see this, recall that all Z-rotation gates are represented by diagonal matrices under the conventional computational basis, also known as the Z-basis. By the same token, all X-rotation gates are diagonal under the X-basis, since we could turn X-rotations into Z-rotations by changing the basis. When the input state of the circuit is $\ket{+}^{\otimes 4}$ as shown in the figure, we observe two different types of symmetries as follows:
\begin{figure}
\centering
\includegraphics[width=.36\textwidth]{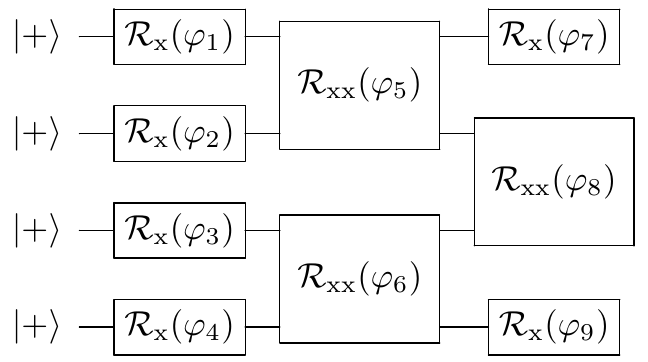}
\vspace{0mm}
\caption{A quantum circuit having symmetries that may be viewed from both state-oriented and circuit-oriented perspectives.}
\label{fig:symmetric_circuit}
\vspace{0mm}\end{figure}
\begin{itemize}
  \item \textbf{State symmetry}: The output state of the circuit has the stabilizer $\mathcal{S}=\M{X}_1\M{X}_2\M{X}_3\M{X}_4$, where $\M{X}_i$ denotes the Pauli-X operator acting on the $i$-th qubit. This stabilizer may be used to detect Z-errors.
  \item \textbf{Circuit symmetry}: This circuit can be diagonalized under the X-basis. Consequently, we have:
  1) Every gate in this circuit commutes with one another; 2) The circuit commutes with the operator $\mathcal{S}$.
\end{itemize}

Observe that in this simple example, the circuit symmetries are more fundamental and more essential than the state symmetry. Indeed, the stabilizer $\mathcal{S}$ originates from the fact that the circuit commutes with $\mathcal{S}$, and that the input state is an eigenstate of $\mathcal{S}$. If the input state is different, the state may no longer be stabilized by $\mathcal{S}$, and hence symmetry verification techniques based on stabilizer checks are no longer applicable. However, the circuit symmetries are still valid in this case. This motivates us to design efficient techniques for verifying circuit symmetries and for mitigating errors that violate these symmetries.\footnote{It is also noteworthy that one may conceive beneficial joint verification schemes of both circuit and state symmetries. For example, one may first encode the quantum state (hence the corresponding circuit) with quantum error correction codes, and then verify the intrinsic symmetries of the encoded circuit for further error mitigation.}

\section{Verifying Gate Commutativity Using Quantum Switch}\label{sec:quantum_switch}
In this section, we show that the commutativity of gates in a quantum circuit could be verified by exploiting the concept of quantum switches. Note that this is a weaker circuit symmetry compared to ``the circuit commutes with some known operator'', which will be investigated in the next section.

Quantum switches constitute a physical realization of the superposition of causal orders, producing quantum states that are coherent superpositions of the outputs of certain quantum circuits. These circuits contain the same operations, but are executed in different sequential orders. Quantum switches have received the attention of both communication and information theorists, since they have been shown to have the potential of improving the overall capacity by superposing certain noisy channels \cite{qs_comm4}. In its simplest form, the quantum switch involving  {a pair of}  channels $\Sop{A}$ and $\Sop{B}$ would effectively produce a superposition of $\Sop{A}\circ \Sop{B}$ and $\Sop{B}\circ \Sop{A}$, with the assistance of a control qubit. The composite channel may be represented as follows \cite{qs_comm4}:
\begin{equation}
\Sop{C}(\rho,\omega) = \sum_{i,j} \M{C}_{ij}(\rho\otimes\omega) \M{C}_{ij}^{\dagger},
\end{equation}
where $\rho$ and $\omega$ represent the state of the data register and the control qubit, respectively, while $\M{C}_{ij}$ denotes a Kraus operator of $\Sop{C}$ given by
\begin{equation}\label{qs_def}
\M{C}_{ij} = \M{A}_i\M{B}_j\otimes \ket{0}\!\bra{0}+\M{B}_j\M{A}_i\otimes \ket{1}\!\bra{1},
\end{equation}
with $\M{A}_i$ and $\M{B}_j$ denoting the Kraus operators of $\Sop{A}$ and $\Sop{B}$, respectively. We observe from \eqref{qs_def} that $\Sop{A}\circ\Sop{B}$ is applied when we measure a $\ket{0}$ on the control qubit, and $\Sop{B}\circ\Sop{A}$ is applied otherwise. This suggest that if the control qubit is set to be a superposition of $\ket{0}$ and $\ket{1}$, the resulting channel would be a superposition of $\Sop{A}\circ\Sop{B}$ and $\Sop{B}\circ\Sop{A}$. A representative example showing the information-theoretic advantage of the quantum switch is that, when both $\Sop{A}$ and $\Sop{B}$ are entanglement-breaking channels (which are extremely noisy) given by \cite{qs_ieee}
\begin{equation}
\Sop{A}(\rho)=\Sop{B}(\rho) = \frac{1}{2}(\M{X}\rho\M{X}+\M{Y}\rho\M{Y}),
\end{equation}
then we obtain a noiseless channel by performing post-selection based on the control qubit.

Inspired by the example of entanglement-breaking channels, we propose to verify the commutativity of gates using quantum switches. Intuitively, we first prepare the control qubit at a superposition state of $\ket{0}$ and $\ket{1}$ in order to produce a superposition of $\Sop{A}\circ \Sop{B}$ and $\Sop{B}\circ \Sop{A}$. Then, conditioned on the measured outcome of the control qubit, we discard the computational results corresponding to the non-commutative components. Formally speaking, we have the following result.
\begin{proposition}\label{prop:qs}
Suppose that the control qubit is initialized to the state $\ket{+}$. If we do not discard any result, the state of the data register is\footnote{The commutator and the anti-commutator between two matrices $\M{A}$ and $\M{B}$ are defined as $[\M{A},\M{B}]:=\M{A}\M{B}-\M{B}\M{A}$ and $\{\M{A},\M{B}\}:=\M{A}\M{B}+\M{B}\M{A}$, respectively.}
\begin{equation}
\rho_{\rm raw}=\sum_{i,j}\frac{\{\M{A}_i,\M{B}_j\}}{2}\rho\frac{\{\M{A}_i,\M{B}_j\}^\dagger}{2}+\frac{[\M{A}_i,\M{B}_j]}{2}\rho\frac{[\M{A}_i,\M{B}_j]^\dagger}{2}.
\end{equation}
By contrast, if we do discard the state once we measure a $\ket{-}$ at the output of the quantum switch, the state of the data register is given by
\begin{equation}
\rho_{\rm out}=\frac{1}{Z}\sum_{i,j}\frac{\{\M{A}_i,\M{B}_j\}}{2}\rho\frac{\{\M{A}_i,\M{B}_j\}^\dagger}{2},
\end{equation}
where $Z$ is a normalization factor given by
$$
Z = \frac{{\rm tr}\{\sum_{i,j}\{\M{A}_i,\M{B}_j\}\rho\{\M{A}_i,\M{B}_j\}^\dagger+[\M{A}_i,\M{B}_j]\rho[\M{A}_i,\M{B}_j]^\dagger\}}{{\rm tr}\{\sum_{i,j}\{\M{A}_i,\M{B}_j\}\rho\{\M{A}_i,\M{B}_j\}^\dagger\}}.
$$
\begin{IEEEproof}
Please refer to Appendix \ref{sec:proof_qs}.
\end{IEEEproof}
\end{proposition}

From Proposition \ref{prop:qs} we see that with the help of the quantum switch, we may filter out the components taking the form of $[\M{A}_i,\M{B}_j]\rho[\M{A}_i,\M{B}_j]^\dagger$ from the output state. Since $\Sop{A}\circ\Sop{B}$ should be equivalent to $\Sop{B}\circ\Sop{A}$ if both $\Sop{A}$ and $\Sop{B}$ are noiseless, we have $[\M{A}_i,\M{B}_j]=0$ under the noise-free condition. This implies that by filtering out components like $[\M{A}_i,\M{B}_j]\rho[\M{A}_i,\M{B}_j]^\dagger$, we may mitigate the computational error. To elaborate further, let us consider the classical average of the computational results of $\Sop{A}\circ\Sop{B}$ and $\Sop{B}\circ\Sop{A}$, which may be expressed as
\begin{equation}
\rho_{\rm avg} = \frac{1}{2} \sum_{i,j}\M{A}_i\M{B}_j\rho\M{B}_j^\dagger\M{A}_i^\dagger + \M{B}_j\M{A}_i\rho\M{A}_i^\dagger\M{B}_j^\dagger.
\end{equation}
After some further manipulations, one would obtain $\rho_{\rm avg}=\rho_{\rm raw}$. This means that by combining a quantum switch and post-selection, we could indeed eliminate certain error components in the raw output state that do not satisfy the gate commutativity conditions.

\subsection{Circuit Implementation and Practical Issues}\label{sec:qs_practical}
\begin{figure}
\centering
\includegraphics[width=.3\textwidth]{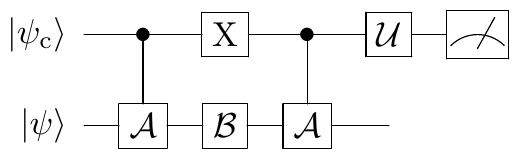}
\vspace{0mm}
\caption{The quantum circuit implementation of a quantum switch between two commuting gates.}
\label{fig:quantum_switch}
\vspace{0mm}
\end{figure}

The quantum switch between two commuting gates $\Sop{A}$ and $\Sop{B}$ can be implemented with the aid of a control qubit \cite{qs_comp}, as portrayed in Fig. \ref{fig:quantum_switch}. The states $\ket{\psi_{\rm c}}$ and $\ket{\psi}$ represent the states of the control qubit and that of the data register, respectively. The gate $\Sop{U}$ is applied for rotating the control qubit so that its state becomes diagonal under the Z-basis. For example, the control qubit is typically initialized to the state $\ket{\psi_{\rm c}}=\ket{+}$, and thus the corresponding $\Sop{U}$ is the Hadamard gate. Upon measuring a $\ket{0}$ on the control qubit, we know that the commutativity between gates $\Sop{A}$ and $\Sop{B}$ is preserved. Otherwise, we discard the computational result. Note that due to the controlled-$\Sop{A}$ gate in Fig.~\ref{fig:quantum_switch}, errors on the data register may also have an effect on the control qubit. For example, let us consider the scenario where an error $\Sop{E}(\rho)=\M{E}\rho\M{E}^\dagger$ that anti-commutes with $\Sop{A}$ is inflicted on the data register before a controlled-$\Sop{A}$ gate, namely we have $\M{E}\M{A}=-\M{A}\M{E}$. Upon denoting the joint state of the control qubit and the data register by $\ket{\varphi}$, we obtain
$$
\M{G}_{\Sop{A}}^{\rm c}(\M{I}\otimes \M{E})\ket{\varphi} = -(\M{I}\otimes \M{E})\M{G}_{\Sop{A}}^{\rm c}\ket{\varphi},
$$
where $\M{G}_{\Sop{A}}^{\rm c}$ denotes the controlled-$\Sop{A}$ gate. This implies that the error $\Sop{E}$ propagates through the controlled-$\Sop{A}$ gate, but additionally it also inflicts a global phase flip, which also affects the control qubit.

\begin{figure*}
\centering
\subfloat[][The implementation relying on multiple control qubits]{
\includegraphics[width=.65\textwidth]{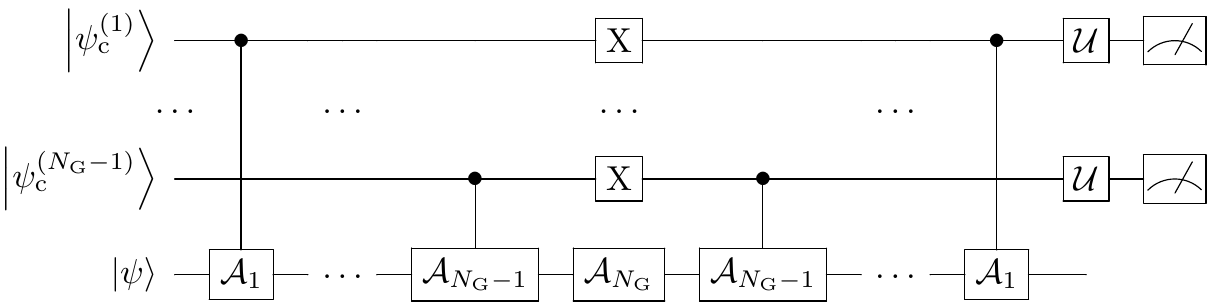}
\label{fig:quantum_switch_multiple_gates}
}\\
\subfloat[][The implementation using a single control qubit]{
\includegraphics[width=.6\textwidth]{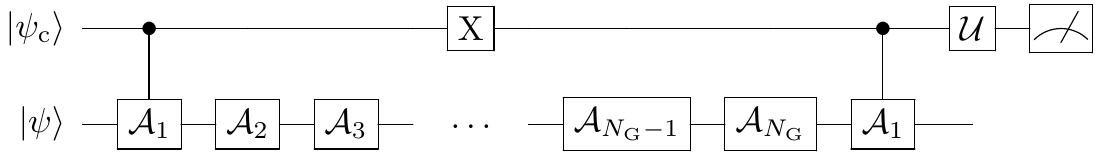}
\label{fig:quantum_switch_multiple_gates2}
}
\caption{Possible generalizations of the quantum switch method to circuits containing $N_{\rm G}>2$ commuting gates.}
\vspace{0mm}
\end{figure*}

There are some noteworthy issues associated with this implementation, when we apply it to practical quantum circuits. First of all, one of the two gates (e.g. the gate $\Sop{A}$ in Fig. \ref{fig:quantum_switch}) has to be implemented in a controlled form, which increases the number of qubits that it acts upon. In practice, a quantum gate acting on more qubits is typically noisier than those acting on less qubits. Therefore, it is not clear whether the quantum switch method achieves a practical accuracy improvement over the original (unprotected) circuit. Another issue is that there is no natural and unified generalization of the method to $N_{\rm G}>2$ gates under the gate model.\footnote{Natural generalizations do exist for other models of quantum computation, for example, photonic quantum computers using the implementation described in \cite{qs_prl}.} Here we present some possible generalizations relying on multiple control qubits, portrayed in Fig. \ref{fig:quantum_switch_multiple_gates} and \ref{fig:quantum_switch_multiple_gates2}.

\section{Verifying the Commutativity with Known Unitaries: Spatio-Temporal Stabilizers}\label{sec:sts}
In the previous section, we have shown that quantum switches could be used to verify the commutativity of quantum gates. But in some practical scenarios, we may have a stronger circuit symmetry, in the sense that a block of gates commute with some known unitaries. For example, in the QAOA, the part implementing a phase Hamiltonian commutes with all Pauli operators containing only Pauli-I and Pauli-Z operators. Intuitively, this stronger sense of symmetry may lead to better error mitigation performance than that of gate commutativity.

\subsection{Improving the Quantum Switch Method}\label{ssec:qs_mod}
In fact, we could verify this strong sense of circuit symmetry by slightly modifying the quantum switch method. Let us denote the circuit to be verified as $\Sop{C}(\rho)=\sum_i \M{C}_i\rho\M{C}_i^\dagger$, and assume that the noiseless component in the circuit, represented by the Kraus operator $\M{C}_1$, commutes with the operator $\Sop{U}(\rho)=\M{U}\rho\M{U}^\dagger$. By applying a quantum switch between $\Sop{C}\circ\Sop{U}$ and $\Sop{U}\circ\Sop{C}$, we obtain the following composite circuit
\begin{equation}
\Sop{D}(\rho,\omega) = \sum_{i} \M{D}_{i}(\rho\otimes\omega) \M{D}_{i}^{\dagger},
\end{equation}
where
\begin{equation}
\M{D}_i = \M{C}_i\M{U}\otimes \ket{0}\!\bra{0}+\M{U}\M{C}_i\otimes \ket{1}\!\bra{1}.
\end{equation}
Similar to the result in Proposition \ref{prop:qs}, after applying $\Sop{D}$, the output state is given by
\begin{equation}
\rho_{\rm m}\propto\sum_{i}\frac{\{\M{C}_i,\M{U}\}}{2}\rho\frac{\{\M{C}_i,\M{U}\}^\dagger}{2}.
\end{equation}
Now we have a coherent superposition of $\Sop{C}\circ\Sop{U}$ and $\Sop{U}\circ\Sop{C}$. But in order to verify the strong circuit symmetry, we do not need to actually apply $\Sop{U}$, which differs from the case discussed in the previous section. In light of this, we apply the inverse of $\Sop{U}$, namely $\Sop{U}^\dagger$, to $\rho_{\rm m}$ and obtain the final output as
\begin{equation}\label{output_sts}
\rho_{\rm out}\propto \sum_i \frac{\M{C}_i+\M{U}^\dagger\M{C}_i\M{U}}{2}\rho\frac{\M{C}_i^\dagger+\M{U}^\dagger\M{C}_i^\dagger\M{U}}{2}.
\end{equation}
In this way, we eliminate the impact of $\Sop{U}$ on the noiseless component in the final result by exploiting the commutativity between $\M{U}$ and $\M{C}_1$. Indeed, observe from \eqref{output_sts} that for the noiseless component $\M{C}_1$, we have
\begin{equation}
\frac{\M{C}_1+\M{U}^\dagger\M{C}_1\M{U}}{2} = \M{C}_1,
\end{equation}
since $\M{U}\M{C}_1=\M{C}_1\M{U}$, implying that it remains unchanged by our modified quantum switch.

We could gain further insights into the error mitigation performance of this modified quantum switch by considering more specific noise models. Observe that each Kraus operator $\M{C}_i$ can be decomposed as $\widetilde{\M{C}}_i\M{C}_1$, representing the noiseless circuit followed by some quantum channel modeling the noise. This follows from the fact that the noiseless circuit $\M{C}_1$ is unitary, hence we can always construct $\widetilde{\M{C}}_i=\M{C}_i\M{C}_1^\dagger$. \footnote{When the error is coherent, there is only one Kraus operator $\M{C}_1$, which is unitary. By contrast, when the error is incoherent, there could be more than one Kraus operators, and these operators may or may not be unitary.} For the noiseless component we have $\widetilde{\M{C}}_1=\M{I}$. Thus we may obtain
\begin{equation}\label{sts_kraus_1}
\begin{aligned}
\M{C}_i+\M{U}^\dagger\M{C}_i\M{U}&=\widetilde{\M{C}}_i\M{C}_1+\M{U}^\dagger\widetilde{\M{C}}_i\M{C}_1\M{U}\\
&=\left(\widetilde{\M{C}}_i+\M{U}^\dagger\widetilde{\M{C}}_i\M{U}\right)\M{C}_1.
\end{aligned}
\end{equation}
Let us assume that the symmetry operator $\M{U}$ is a Pauli operator, which is common for practical quantum circuits. Note that among the group of Pauli operators, given a fixed operator $\M{U}$, any other operator either commutes with $\M{U}$ or anti-commutes with $\M{U}$. This implies that $\widetilde{\M{C}}_i$ may be decomposed into two parts as
\begin{equation}
\widetilde{\M{C}}_i = \widetilde{\M{C}}_i^{(\rm c)}+\widetilde{\M{C}}_i^{(\rm a)},
\end{equation}
where $\widetilde{\M{C}}_i^{(\rm c)}$ commutes with $\M{U}$ and $\widetilde{\M{C}}_i^{(\rm a)}$ anti-commutes with $\M{U}$. This is because all quantum operations can be represented as linear combinations of Pauli operators. Therefore, \eqref{sts_kraus_1} can be further simplified as
\begin{equation}
\begin{aligned}
\M{C}_i+\M{U}^\dagger\M{C}_i\M{U} &= \Big(\widetilde{\M{C}}_i^{(\rm c)}+ \widetilde{\M{C}}_i^{(\rm a)}\Big)\M{C}_1 +\Big(\widetilde{\M{C}}_i^{(\rm c)}- \widetilde{\M{C}}_i^{(\rm a)}\Big)\M{C}_1\\
&= 2\widetilde{\M{C}}_i^{(\rm c)}\M{C}_1,
\end{aligned}
\end{equation}
since
$$
\M{U}^\dagger\widetilde{\M{C}}_i^{(\rm a)}\M{U} = -\widetilde{\M{C}}_i^{(\rm a)},~
\M{U}^\dagger\widetilde{\M{C}}_i^{(\rm c)}\M{U}=\widetilde{\M{C}}_i^{(\rm c)}.
$$
Hence we have
\begin{equation}
\rho_{\rm out}\propto \sum_i \widetilde{\M{C}}_i^{(\rm c)}\M{C}_1\rho\M{C}_1^\dagger\Big(\widetilde{\M{C}}_i^{(\rm c)}\Big)^\dagger.
\end{equation}

One could verify that similar arguments can also be applied to the case where $\Sop{C}$ consists of more than one noisy gates. For example, when there are two noisy gates in the circuit, the Kraus operators satisfy $\M{C}_{ij} = \widetilde{\M{C}}_{1,i}\M{C}_{1,1}\widetilde{\M{C}}_{2,j}\M{C}_{2,1}$, and we have
\begin{equation}\label{sts_general}
\begin{aligned}
&\M{C}_{ij}+\M{U}^\dagger\M{C}_{ij}\M{U} \\
&\hspace{3mm}= \widetilde{\M{C}}_{1,i}^{(\rm c)}\M{C}_{1,1}\widetilde{\M{C}}_{2,i}^{(\rm c)}\M{C}_{2,1} + \widetilde{\M{C}}_{1,i}^{(\rm a)}\M{C}_{1,1}\widetilde{\M{C}}_{2,i}^{(\rm a)}\M{C}_{2,1},
\end{aligned}
\end{equation}
as long as both $\M{C}_{1,1}$ and $\M{C}_{2,1}$ commute with $\M{U}$. We may infer from \eqref{sts_general} that:
\begin{remark}
Let us consider the scenario where the channels of each gate only impose anti-commutative errors (e.g. bit-flip channels when $\M{U}=\M{Z}$). When the anti-commutative error operators such as $\widetilde{\M{C}}_{1,i}^{(\rm a)}$ (and also others with different subscripts) satisfy $\|\widetilde{\M{C}}_{1,i}^{(\rm a)}\|=O(\sqrt{\epsilon})$ where $\epsilon$ denotes the average error rate per gate, upon the verification of the commutativity with $\Sop{U}$, the residual error rate for a circuit containing multiple noisy gates is on the order of $O(\epsilon^2)$. To elaborate, any error pattern constituted by an odd number of anti-commutative error operators would be mitigated, hence the dominant residual error patterns would incorporate at least two anti-commutative error operators, taking the following form:
\begin{equation}
\widetilde{\M{C}}_{\rm res} = \widetilde{\M{C}}_{m,i}^{(\rm a)}\widetilde{\M{C}}_{n,j}^{(\rm a)}\rho (\widetilde{\M{C}}_{n,j}^{(\rm a)})^\dagger (\widetilde{\M{C}}_{m,i}^{(\rm a)})^\dagger,
\end{equation}
which is on the order of $O(\epsilon^2)$.

\end{remark}

In particular, when the error operators are Pauli operators, we have the following explicit result:
\begin{remark}
In general, any Pauli operator constituted by the tensor product of an even number of Pauli-Zs would commute with $\M{X}^{\otimes N}$, whereas it would anti-commute with $\M{X}^{\otimes N}$, if the number of Pauli-Z's is odd.
\end{remark}

Intuitively, when the errors act independently upon each qubit, by verifying a circuit symmetry $\Sop{U}$ which is a Pauli operator, we may detect any single-qubit anti-commutative error. This resembles the effect of error-detecting stabilizer codes. Partly for this reason, we will refer to the aforementioned modified quantum switch method as the \emph{spatio-temporal stabilizer} method in the rest of this treatise. This terminology will be explained in more detail in Section \ref{ssec:sts}.

\subsection{Implementation: Spatio-temporal Stabilizer Check}\label{ssec:sts}
\begin{figure}
\centering
\subfloat[][Direct implementation]{
\includegraphics[width=.24\textwidth]{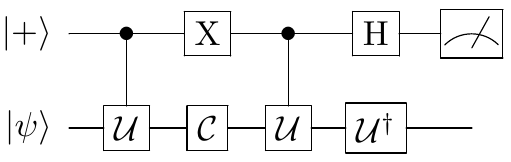}
\label{fig:sts_circuit}
}
\subfloat[][Simplified implementation]{
\includegraphics[width=.24\textwidth]{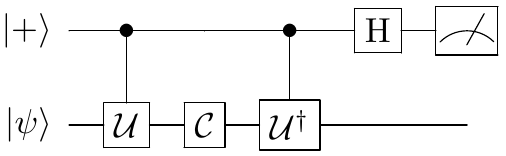}
\label{fig:sts_circuit_simplified}
}
\caption{Circuit implementations of an \ac{sts} check.}
\vspace{0mm}
\end{figure}
According to the discussion in Section \ref{ssec:qs_mod}, we could readily obtain a circuit implementing the modified quantum switch portrayed in Fig. \ref{fig:sts_circuit}. But this circuit admits a simplification, as portrayed in Fig. \ref{fig:sts_circuit_simplified}, which helps us better understand this method. As it may be observed from the figure, the final state of the data register would be $\Sop{U}^\dagger\circ\Sop{C}\circ\Sop{U}(\ket{\psi}\!\bra{\psi})$ if the control qubit is in $\ket{1}$, when the controlled-$\Sop{U}$ and controlled-$\Sop{U}^\dagger$ gates are being applied, and $\Sop{C}(\ket{\psi}\!\bra{\psi})$ if the control qubit is in $\ket{0}$. But the control qubit is in $\ket{+}$ due to the Hadamard gate, hence if we measure a $\ket{0}$ on the control qubit at the output of the circuit, the Kraus operators on the data register are given by
\begin{equation}
\M{K}_i = \frac{1}{2}\left(\M{C}_i+\M{U}^\dagger\M{C}_i\M{U}\right),
\end{equation}
as we have expected.

\begin{figure}
\centering
\subfloat[][The \ac{sts} $\Sop{S}\{\M{X}_1(0),\M{X}_2(0),\M{X}_1(1),\M{X}_2(1)\}$]{
\includegraphics[width=.35\textwidth]{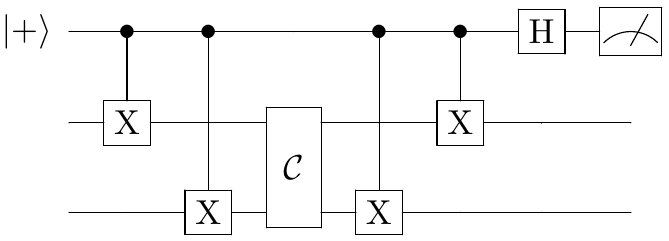}
\label{fig:sts_measure_pr}
}
\hspace{5mm}
\subfloat[][The conventional stabilizer $\M{X}_1\M{X}_2$]{
\includegraphics[width=.24\textwidth]{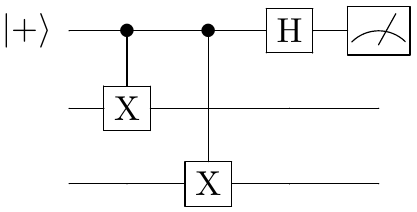}
\label{fig:stabilizer_measure_pr}
}
\caption{Comparison between the circuit implementation of \acp{sts} and that of conventional stabilizers.}
\vspace{0mm}
\end{figure}

To gain further intuition, we consider a toy example, where the circuit $\Sop{C}$ commutes with a Pauli operator $\Sop{U}$ given by $\M{U}=\M{X}_1 \M{X}_2$. In this case, the simplified circuit can be constructed as shown in Fig. \ref{fig:sts_measure_pr}. From this figure we see that the simplified circuit is rather similar to the ones performing stabilizer checks. For example, if we wish to measure a stabilizer $\M{X}_1\M{X}_2$, we could use the circuit portrayed in Fig. \ref{fig:stabilizer_measure_pr}. Compared to Fig. \ref{fig:stabilizer_measure_pr}, the circuit in Fig. \ref{fig:sts_measure_pr} looks like measuring a stabilizer in a bipartite manner, for which a part is applied before the circuit $\Sop{C}$, and the rest of it is applied after $\Sop{C}$. In fact, upon denoting the input state of the data register as $\ket{\psi}$, it is clear that the output state $\M{C}\ket{\psi}$ has the following stabilizer
\begin{equation}
\M{S} = \M{C}(\M{X}_1 \M{X}_2)\M{C}^\dagger(\M{X}_1 \M{X}_2),
\end{equation}
and that the circuit in Fig. \ref{fig:sts_measure_pr} indeed measures the stabilizer $\M{S}$. Since the gates in quantum circuits are executed in a sequential manner, if we define the time right before $\Sop{C}$ is applied as $t=0$, and the time right after $\Sop{C}$ is applied as $t=1$, we see that the stabilizer $\M{S}$ contains a $(\M{X}_1\M{X}_2)$ at time $t=0$, and another $(\M{X}_1\M{X}_2)$ at time $t=1$. Therefore, we refer to $\M{S}$ as a ``\emph{spatio-temporal stabilizer}'' of the output state $\M{C}\ket{\psi}$, which can be formally defined as follows.

\begin{definition}[Spatio-temporal stabilizer of a state]
Consider a quantum circuit consisting of $N$ unitary gates given by $\M{C} = \M{C}_N\M{C}_{N-1}\dotsc\M{C}_1$, with input state $\ket{\psi}$. We say that $\M{S}$ is a $(N+1)$-partite spatio-temporal stabilizer (STS) of the output state $\M{C}\ket{\psi}$, if it satisfies $\M{S}\M{C}\ket{\psi}=\M{C}\ket{\psi}$, and takes the following form
\begin{equation}\label{sts_def}
\begin{aligned}
\M{S}&=\Sop{S}\{\M{S}_0(0),\M{S}_1(1),\dotsc,\M{S}_N(N)\}\\
&:=\M{C} \M{S}_0^\dagger\M{C}_1^\dagger \M{S}_1^\dagger\dotsc\M{C}_N^\dagger\M{S}_N^\dagger.
\end{aligned}
\end{equation}
The argument $t$ in $\M{S}_n(t)$ represents the time instance when this partial operator is applied. The partial operators $\M{S}_n,~n=1\dotsc N$ are called the \emph{components} of $\M{S}$. When a component $\M{S}_n(t)$ can be represented as the product of several mutually commutative operators, these operators may be viewed as being applied at the same time instance $t$. For example, $\M{S}=\M{X}_1\M{X}_2$ applied at time instance $1$ may be decomposed into $\M{X}_1(1)$ and $\M{X}_2(1)$. In the context of \acp{sts}, we refer to the control qubits as \emph{ancillas} to be consistent with the terminologies in the conventional stabilizer formalism.
\end{definition}

\begin{remark}
The definition \eqref{sts_def} is inspired by the following natural condition
\begin{equation}
\M{S}_N\M{C}_N\M{S}_{N-1}\dotsc \M{C}_1\M{S}_0\ket{\psi} = \M{C}\ket{\psi}
\end{equation}
for the components $\{\M{S}_n\}_{n=0}^N$ to form an \ac{sts}. If we require $\M{S}\M{C}\ket{\psi}=\M{C}\ket{\psi}$ to be satisfied, we have
\begin{equation}\label{sts_condition}
\M{S}_N\M{C}_N\M{S}_{N-1}\dotsc \M{C}_1\M{S}_0\ket{\psi} = \M{S}\M{C}\ket{\psi}.
\end{equation}
If \eqref{sts_condition} is satisfied for all $\ket{\psi}$, the following holds
\begin{equation}
\M{S}_N\M{C}_N\M{S}_{N-1}\dotsc \M{C}_1\M{S}_0\ket{\psi} = \M{S}\M{C},
\end{equation}
and hence we arrive at \eqref{sts_def}.
\end{remark}

When $\M{S}\M{C}\ket{\psi}=\M{C}\ket{\psi}$ is satisfied regardless of the state $\ket{\psi}$, $\M{S}$ may be viewed as a stabilizer of the circuit $\M{C}$ itself, defined as follows.
\begin{definition}[\ac{sts} of a circuit]
We say that $\M{S}$ is a $(N+1)$-partite \ac{sts} \emph{of the circuit} $\M{C}$, if it satisfies $\M{S}\M{C}=\M{C}$, and hence the partial operators $\{\M{S}_n\}_{n=0}^N$ satisfy
\begin{equation}\label{strong_sts}
\M{S}_N\M{C}_N\M{S}_{N-1}\dotsc \M{C}_1\M{S}_0 = \M{C}.
\end{equation}
In the rest of this treatise, if not stated otherwise, all \acp{sts} refer to the \acp{sts} of circuits.
\end{definition}

\begin{figure}[t]
\centering
\includegraphics[width=.4\textwidth]{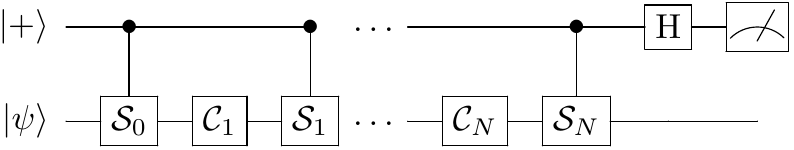}
\vspace{0mm}
\caption{The circuit measuring the \ac{sts} in \eqref{sts_def}.}
\label{fig:sts_general}
\vspace{0mm}\end{figure}

\begin{figure}[t]
\centering
\includegraphics[width=.38\textwidth]{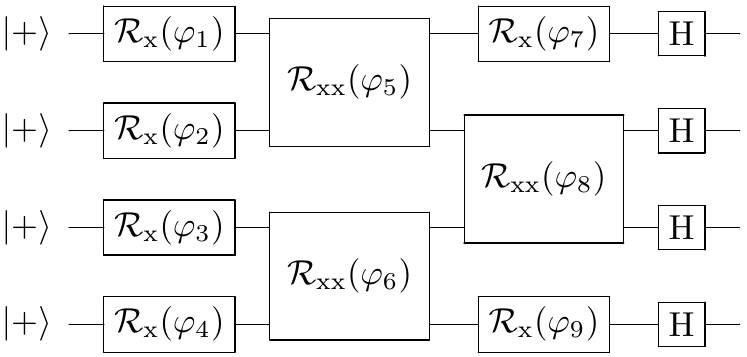}
\vspace{0mm}
\caption{A circuit having an \ac{sts} as in \eqref{sts_example}, but it is difficult to find an operator that commutes with it.}
\label{fig:symmetric_circuit2}
\vspace{0mm}\end{figure}

The circuit measuring the \ac{sts} in \eqref{sts_def} is portrayed in Fig. \ref{fig:sts_general}. We may observe from Fig. \ref{fig:sts_general} that the concept of \ac{sts} actually generalizes the idea of verifying circuit commutativity with known operators, since the partial operators $\Sop{S}_0$ through $\Sop{S}_N$ can all be different. A natural question that arises is, whether this generalization has any practical implication. In fact, we could illustrate the usefulness of this generalization, by revisiting the example in Fig. \ref{fig:symmetric_circuit}. We now see that the circuit commutes with $\M{X}^{\otimes 4}$, and equivalently, we say that the circuit has the \ac{sts}
\begin{equation}
\begin{aligned}
&\Sop{S}^{(\rm st)}\{\M{X}_1(0),\M{X}_2(0),\M{X}_3(0),\M{X}_4(0), \\
&\hspace{10mm}\M{X}_1(1),\M{X}_2(1),\M{X}_3(1),\M{X}_4(1)\}.
\end{aligned}
\end{equation}
But if we further apply a Hadamard gate to each of the qubits at the output of the circuit, as portrayed in Fig. \ref{fig:symmetric_circuit2}, it becomes difficult to find an operator that commutes with the new circuit. By contrast, we could say that this circuit has a different \ac{sts} given by
\begin{equation}\label{sts_example}
\begin{aligned}
\M{S}&=\Sop{S}^{(\rm st)}\{\M{X}_1(0),\M{X}_2(0),\M{X}_3(0),\M{X}_4(0), \\
&\hspace{10mm}\M{Z}_1(1),\M{Z}_2(1),\M{Z}_3(1),\M{Z}_4(1)\},
\end{aligned}
\end{equation}
since the circuit (denoted by $\M{C}$) satisfies
\begin{equation}
\M{Z}^{\otimes 4}\M{C} =\M{C}\M{X}^{\otimes 4}.
\end{equation}

\subsection{Simultaneous Observability of \acp{sts}}\label{ssec:simutaneous_observability}
When we consider the verification of a quantum state or a circuit that has multiple symmetries, a natural requirement is that these symmetries can be checked at the same time. Otherwise, only a subset of the symmetries can be verified in each computation, which may result in an unsatisfactory error mitigation performance.

Simultaneous observability is a natural property of conventional stabilizers \cite[Sec. 10.5.4]{ncbook}. A fundamental characteristic of quantum mechanics is the uncertainty principle, stating that a pair of observables can be simultaneously determined to an arbitrary accuracy, if and only if they commute with each other. Stabilizers, being special cases of observables, also follow this principle. In fact, all stabilizers of the same quantum state commute with one another, and hence they form the so-called stabilizer group \cite[Sec. 10.5.4]{ncbook}. This is easily seen by observing that
\begin{equation}
\begin{aligned}
&\M{S}_1 \ket{\psi}=\ket{\psi}~{\rm AND}~\M{S}_2 \ket{\psi}=\ket{\psi} \\
&\hspace{5mm}\Longrightarrow \M{S}_1\M{S}_2\ket{\psi} = \M{S}_2\M{S}_1 \ket{\psi} =\ket{\psi}.
\end{aligned}
\end{equation}
Therefore, conventional stabilizers of the same state are always simultaneously observable.

\begin{figure}[t]
\centering
\includegraphics[width=.45\textwidth]{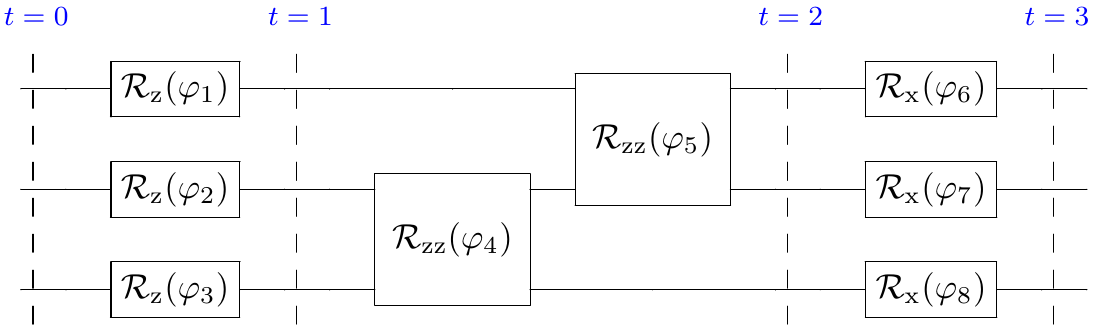}
\vspace{0mm}
\caption{A circuit having two \acp{sts} $\M{S}_1=\Sop{S}\{\M{Z}^{\otimes 3}(0),\M{Z}^{\otimes 3}(2)\}$ and $\M{S}_2=\Sop{S}\{\M{X}^{\otimes 3}(1),\M{X}^{\otimes 3}(3)\}$ that are not simultaneously observable.}
\label{fig:simultaneous_observability}
\vspace{0mm}
\end{figure}

For \acp{sts}, however, simultaneous observability is not necessarily satisfied. To be more specific, let us consider the example portrayed in Fig. \ref{fig:simultaneous_observability}. It is clear that the circuit has two \acp{sts}, namely $\M{S}_1=\Sop{S}\{\M{Z}^{\otimes 3}(0),\M{Z}^{\otimes 3}(2)\}$ and $\M{S}_2=\Sop{S}\{\M{X}^{\otimes 3}(1),\M{X}^{\otimes 3}(3)\}$. However, $\M{S}_1$ and $\M{S}_2$ are not simultaneously observable, since $\M{X}^{\otimes 3}$ does not commute with $\M{Z}^{\otimes 3}$, and hence the combination of $\M{S}_1$ and $\M{S}_2$ given by $\Sop{S}\{\M{Z}^{\otimes 3}(0),\M{X}^{\otimes 3}(1),\M{Z}^{\otimes 3}(2),\M{X}^{\otimes 3}(3)\}$ is not an \ac{sts} of the original circuit. Therefore, we are motivated to propose the following formal definition of simultaneous observability for \acp{sts}.

\begin{definition}[Simultaneous Observability]\label{def:simu}
Consider a set of \ac{sts} checks of a certain circuit $\Sop{C}$, implemented in the fashion shown in Fig. \ref{fig:sts_general} with the aid of ancillas. If the state of the data register at the output of $\Sop{C}$ is the same regardless of the initial states of the ancillas, we say that the \acp{sts} are simultaneously observable.
\end{definition}

Intuitively, by initializing some ancillas to the state $\ket{0}$, we effectively disable certain \acp{sts}. Hence, simultaneous observability means that an arbitrary combination of the \acp{sts} still constitutes an \ac{sts} of the circuit. Unfortunately, determining the simultaneous observability directly using the definition may be inconvenient when the number of \acp{sts} is large, given the excessive number of possible \acp{sts} combinations. To this end, we provide some sufficient conditions that may be useful in practice, based on the following definition of the \textit{action scope} of \acp{sts}.

\begin{figure}[t]
\centering
\subfloat[][Temporally disjoint \acp{sts}.]{
\begin{overpic}[width=.4\textwidth]{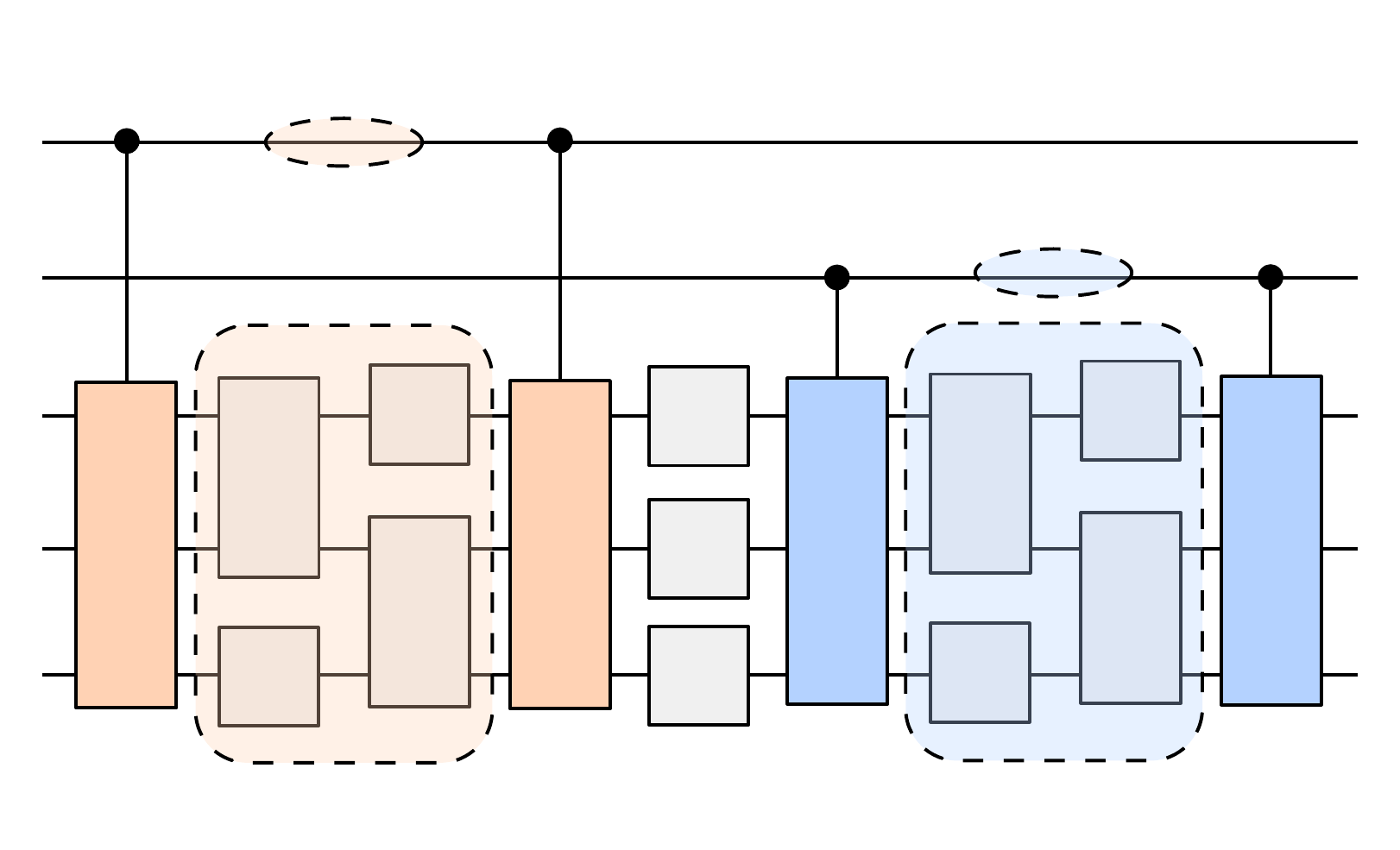}
\put(5,54){\footnotesize \color{red} the control qubit for $\M{S}_1$}
\put(55,44){\footnotesize \color{blue} the control qubit for $\M{S}_2$}
\put(5,1){\footnotesize \color{red} the action scope of $\M{S}_1$}
\put(56,1){\footnotesize \color{blue} the action scope of $\M{S}_2$}
\end{overpic}
\label{fig:disjoint_1}
}
\\
\subfloat[][Spatially disjoint \acp{sts}.]{
\begin{overpic}[width=.4\textwidth]{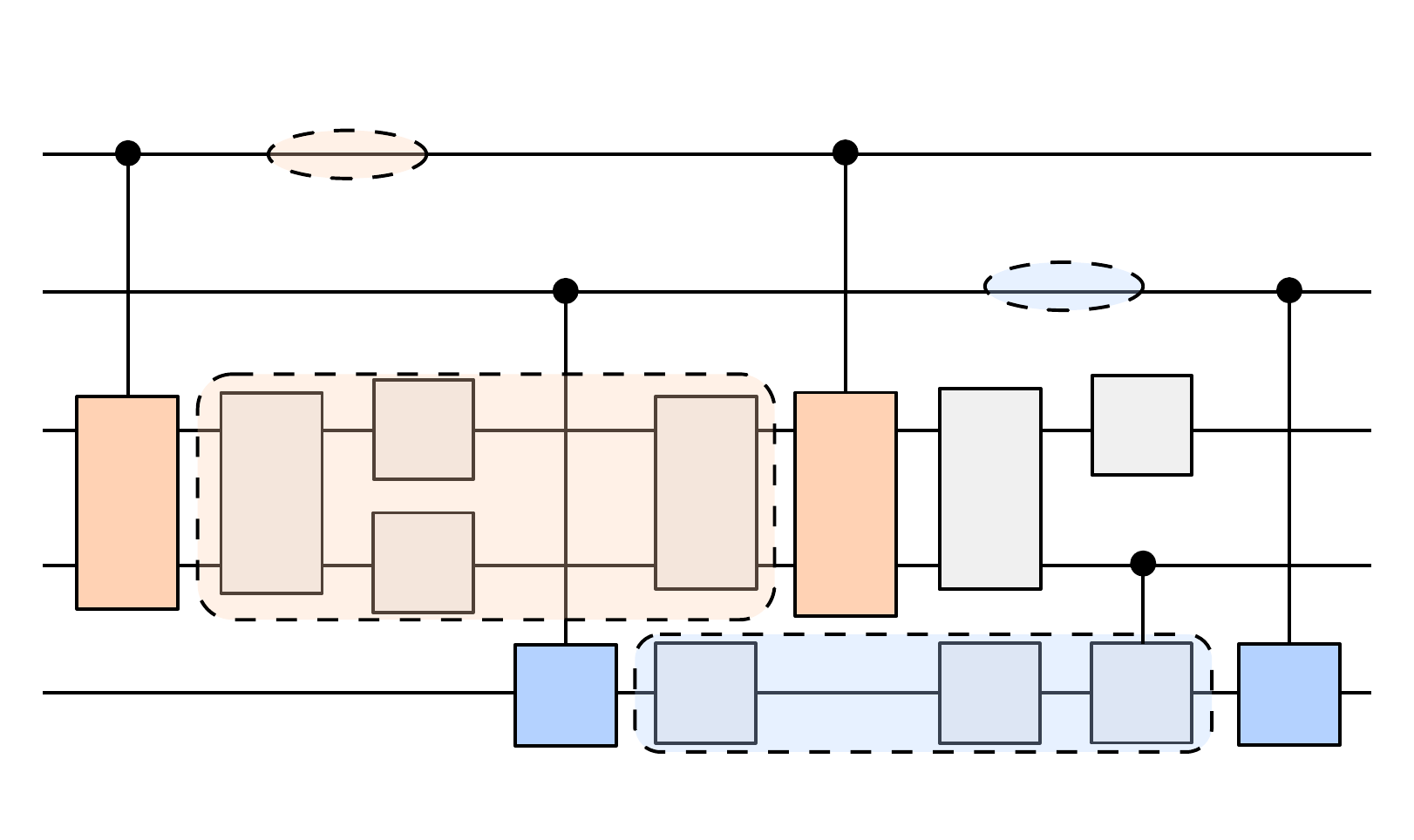}
\put(5,52.5){\footnotesize \color{red} the control qubit for $\M{S}_1$}
\put(55,43){\footnotesize \color{blue} the control qubit for $\M{S}_2$}
\put(15,35){\footnotesize \color{red} the action scope of $\M{S}_1$}
\put(46,2){\footnotesize \color{blue} the action scope of $\M{S}_2$}
\end{overpic}
\label{fig:disjoint_2}
}
\caption{\acp{sts} having disjoint action scopes are simultaneously observable.}
\vspace{0mm}\end{figure}
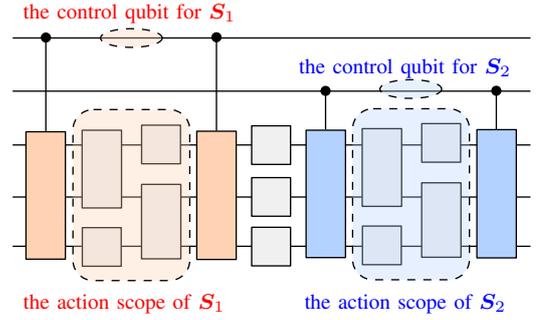
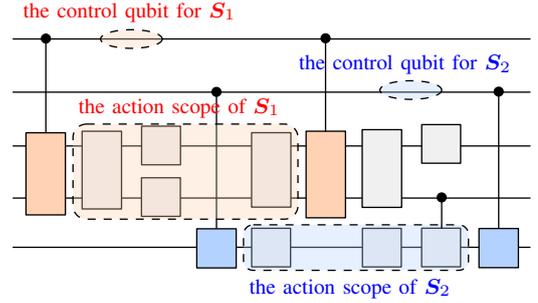

\begin{definition}[Action Scopes]
The action scope of an \ac{sts} $\M{S}$ is a set $\Set{S}=\Set{S}_{\rm s}\times \Set{S}_{\rm t}$, where $\Set{S}_{\rm s}$ is the spatial action scope constituted by the indices of all qubits that the component operators of $\M{S}$ act upon, while $\Set{S}_{\rm t}=\{t|t\le t_{\max},~t\ge t_{\min},~t\in\mathbb{Z}\}$ is the temporal action scope, with $t_{\max}$ and $t_{\min}$ denoting the maximum and the minimum temporal indices in $\M{S}$, respectively.
\end{definition}

To elaborate, for example, the action scope of the \ac{sts} $\M{S}=\Sop{S}\{\M{X}_0(0),\M{Z}_1(0),\M{X}_1(2),\M{Z}_2(3)\}$ is $\{0,1,2\}\times \{0,1,2,3\}$. By exploiting the concept of action scope, the following sufficient condition of simultaneous observability may be obtained.

\begin{sufcondition}[Disjoint Action Scopes]\label{sufcond:disjoint}
If the action scopes of a set of \acp{sts} are mutually disjoint, these \acp{sts} are simultaneously observable.
\begin{IEEEproof}
If the \acp{sts} $\M{S}_1$ and $\M{S}_2$ have disjoint action scopes, they can be viewed as \acp{sts} of two disjoint sub-circuits of the original circuit, respectively, as portrayed in Fig. \ref{fig:disjoint_1}. Hence they are simultaneously observable.
\end{IEEEproof}
\end{sufcondition}

A more sophisticated (and potentially more useful) sufficient condition may be obtained by modifying Sufficient Condition \ref{sufcond:disjoint}, detailed as follows.

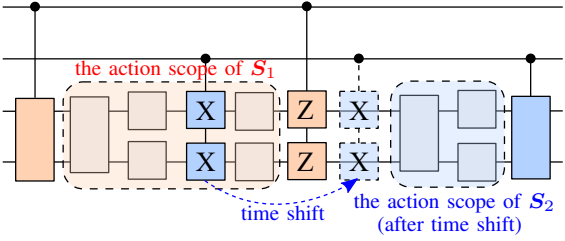
\begin{figure}[t]
\centering
\begin{overpic}[width=.43\textwidth]{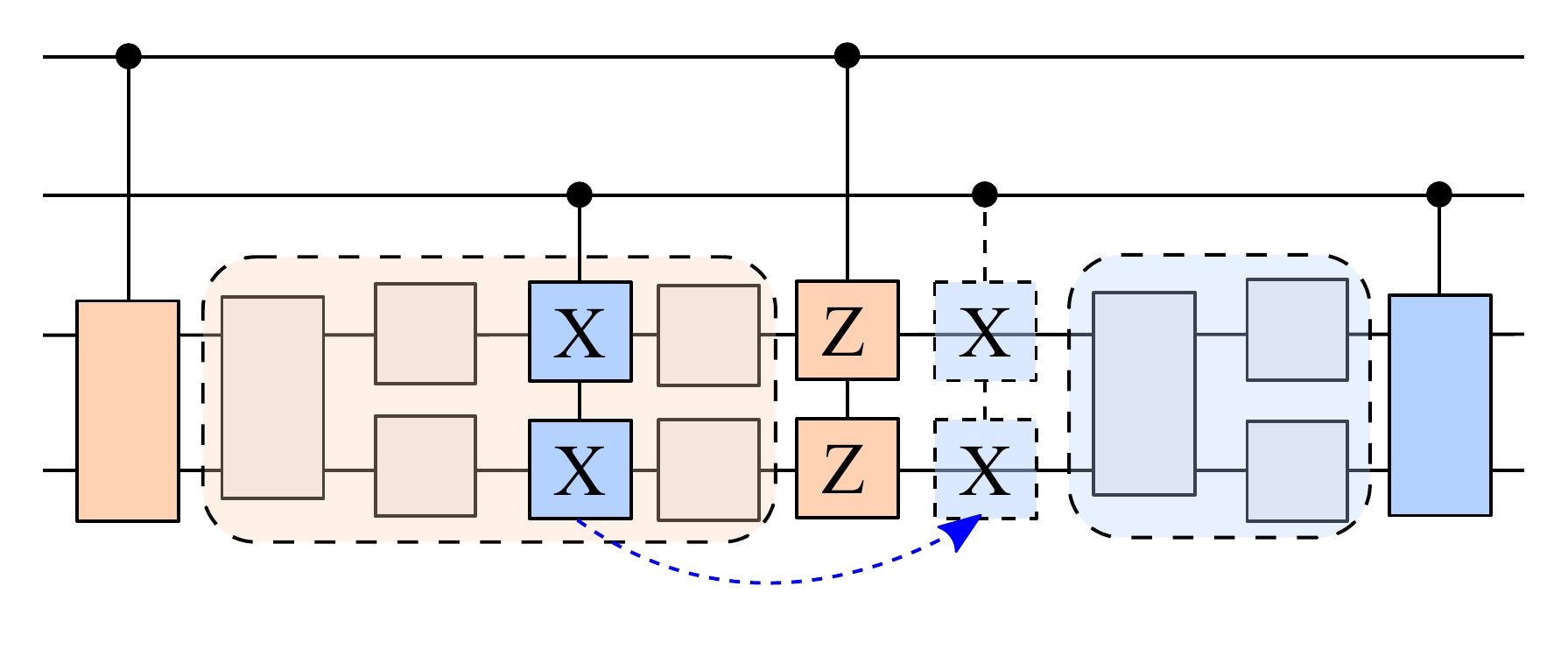}
\put(15,27.5){\footnotesize \color{red} the action scope of $\M{S}_1$}
\put(62,5){\footnotesize \color{blue} the action scope of $\M{S}_2$}
\put(66,1){\footnotesize \color{blue} (after time shift)}
\put(43,3){\footnotesize \color{blue} time shift}
\end{overpic}
\vspace{0mm}
\caption{\acp{sts} having disjoint action scopes after appropriate time shifts are simultaneously observable.}
\label{fig:disjoint_timeshift}
\vspace{0mm}\end{figure}

\begin{sufcondition}[Disjoint Action Scopes After Time Shift]
Consider a set of \acp{sts} $\Set{A}$. The \acp{sts} in $\Set{A}$ are simultaneously observable, if for each $\M{S}_i\in\Set{A}$, we may impose appropriate time shifts to $\forall \M{S}_j\in\Set{A},~j\neq i$, ensuring that the results after the time shifts are still \acp{sts} of the original circuit, and that their action scopes are disjoint with that of $\M{S}_i$. A legitimate time shift for \ac{sts} $\M{S}_j$ is a translation of certain components in $\M{S}_j$ to another time instance, satisfying the condition that these components commute with all the components of other \acp{sts} in $\Set{A}$ lying on the trajectory of the translation, as portrayed in Fig. \ref{fig:disjoint_timeshift}.
\begin{IEEEproof}
Denote the result of time shift for $\M{S}_j$ as $\Sop{T}(\M{S}_j)$. From Sufficient Condition \ref{sufcond:disjoint} we see that $\Sop{T}(\M{S}_j)$ and $\M{S}_i$ are simultaneously observable, and thus the combination of $\M{S}_i$ and $\Sop{T}(\M{S}_j)$ is an \ac{sts}. Since the translated components of $\M{S}_j$ commute with those of other \acp{sts} on the translation trajectory, we see that the combination of $\M{S}_i$ and $\M{S}_j$ is also an \ac{sts}. By applying the arguments to all pairs of \acp{sts} in $\Set{A}$, we arrive at the desired result.
\end{IEEEproof}
\end{sufcondition}

In the example shown in Fig. \ref{fig:disjoint_timeshift}, the \acp{sts} $\M{S}_1$ and $\M{S}_2$ are simultaneously observable, because $\M{X}^{\otimes 2}$ commutes with $\M{Z}^{\otimes 2}$. We will see how this is related to the \acp{sts} of the \ac{qaoa} in Section \ref{ssec:qaoa_sts}.

\subsection{The Accuracy vs. Overhead Trade-off}\label{ssec:accuracy_vs_overhead}
\begin{figure*}[t]
\centering
\begin{overpic}[width=.8\textwidth]{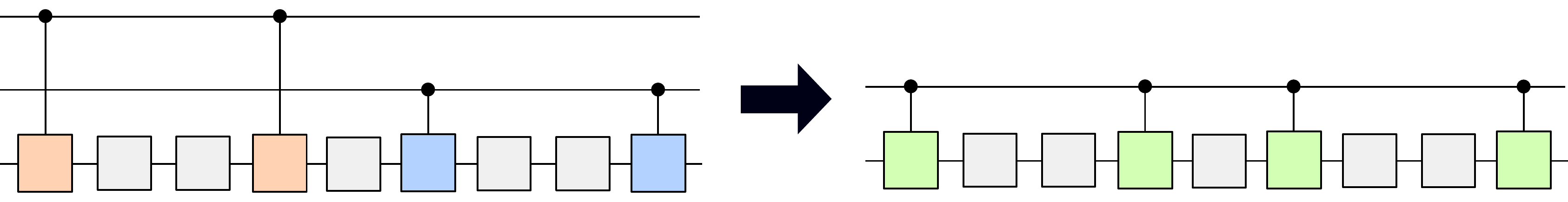}
\end{overpic}
\vspace{0mm}
\caption{Reducing the overhead of control qubits by combining simultaneously observable \acp{sts}.}
\label{fig:combine_sts}
\vspace{0mm}
\end{figure*}

According to the discussion in Section \ref{ssec:sts}, by default, we use one ancilla for checking each \ac{sts}. In fact, we could reallocate the qubit resources exploited for controlling \acp{sts} to strike more flexible accuracy vs. overhead trade-offs. For example, we may combine several simultaneously observable \acp{sts} into a single \ac{sts} to reduce the overall qubit overhead, as portrayed in Fig. \ref{fig:combine_sts}.

\begin{figure*}[t]
\centering
\begin{overpic}[width=.7\textwidth]{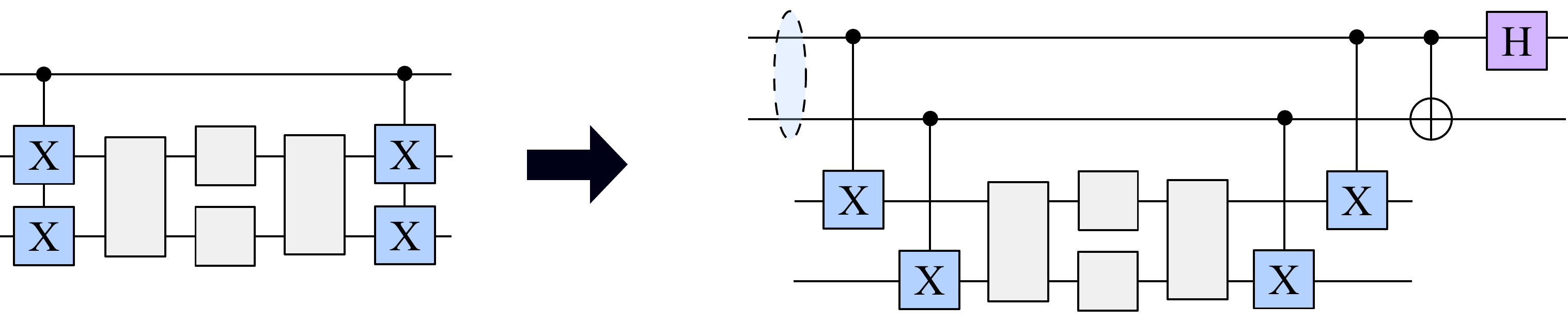}
\put(37.5,14.75){\color{blue} $\frac{\ket{00}+\ket{11}}{\sqrt{2}}$}
\end{overpic}
\vspace{0mm}
\caption{Mitigating error proliferation by measuring a single \ac{sts} relying on multiple control qubits.}
\label{fig:split_sts}
\vspace{0mm}
\end{figure*}

The overhead reduction obtained by combining \acp{sts} comes at a price of stronger error proliferation. To elaborate, observe that  in the circuits shown in Fig. \ref{fig:combine_sts}, the errors may propagate from the ancillas to the data register. However, the circuit on the right hand side suffers from more severe error proliferation, since the errors in the data register may propagate to the control, and then back to the data register. Therefore, when a higher accuracy is required and the qubit resources are abundant, we may measure a single \ac{sts} using multiple ancillas to mitigate error proliferation, relying on pre-shared entanglements between the ancillas (i.e., the ``cat'' state \cite[Sec. 10.6.3]{ncbook}), as portrayed in Fig. \ref{fig:split_sts}. This implementation bears some similarity with the fault-tolerant measurements of conventional stabilizers \cite[Sec. 10.6.3]{ncbook}.

Another type of computational overhead is the sampling overhead, which originates from the fact that some computational results are discarded due to their failure to pass the \ac{sts} checks. To quantify the sampling overhead, we introduce the concept of sampling overhead factor, originally defined in \cite{sof_analysis} for the analysis of channel inversion-based \ac{qem}.

\begin{definition}[Sampling Overhead Factor]
The sampling overhead factor of a set $\Set{A}$ of \acp{sts} applied to a circuit $\Sop{C}$ is defined as
\begin{equation}\label{def_sof}
{\rm SOF}(\Sop{C},\Set{A}) = \frac{1}{p_{\rm pass}(\Sop{C},\Set{A})}-1,
\end{equation}
where $p_{\rm pass}(\Sop{C},\Set{A})$ denotes the probability that the circuit passes all the \ac{sts} checks in $\Set{A}$.
\end{definition}

We will characterize the sampling overhead factors of the \acp{sts} applied to some practical quantum circuits in Section \ref{sec:numerical}.

\section{Case Study: The \acp{sts} of the \ac{qft} and the \ac{qaoa}}\label{sec:case_study}
In this section, we demonstrate the applicability and the characteristics of the \ac{sts} method using two classes of practical quantum circuits, namely that of the \ac{qft} and the \ac{qaoa}.

\subsection{The \acp{sts} of the \ac{qft} Circuits}
The \ac{qft} serves as a subroutine in the quantum phase estimation algorithm, which in turn plays significant roles in other more sophisticated quantum algorithms, including Shor's algorithm and the Harrow-Hassidim-Lloyd (HHL) algorithm \cite{shor,hhl}. Therefore, mitigating the error in the \ac{qft} is beneficial for a range of quantum algorithms.

\begin{figure*}[t]
\centering
\begin{overpic}[width=.7\textwidth]{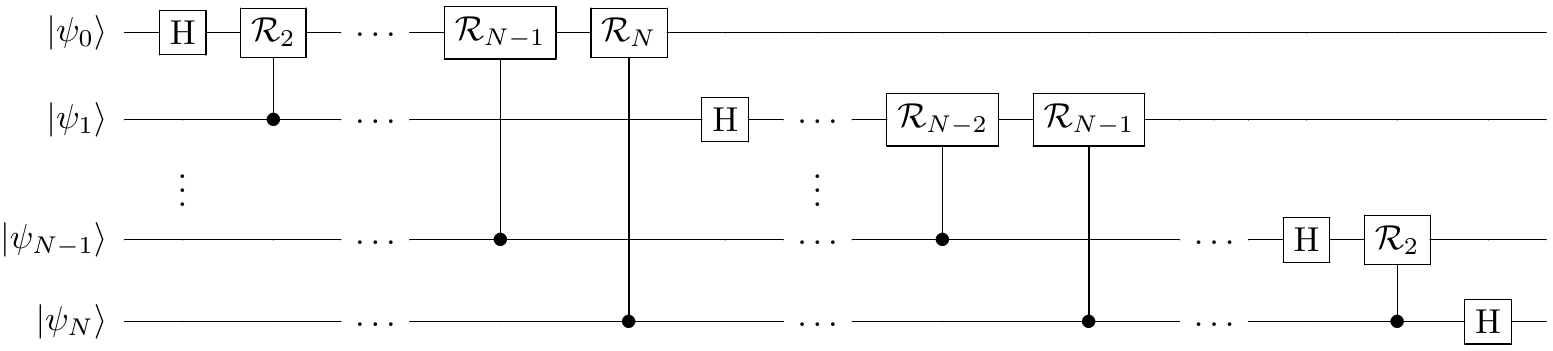}
\end{overpic}
\vspace{0mm}
\caption{The circuit implementing an $N$-qubit \ac{qft}.}
\label{fig:sts_qft}
\end{figure*}

The structure of an $N$-qubit \ac{qft} circuit is portrayed in Fig. \ref{fig:sts_qft}, where the operator $\Sop{R}_n$ (in the controlled-$\Sop{R}_n$ gates) is a single-qubit Z-rotation defined by
\begin{equation}
\Sop{R}_n=\ket{0}\!\bra{0}+e^{\imath 2\pi  2^{-n}} \ket{1}\!\bra{1}.
\end{equation}
It is clearly seen from the figure that each qubit in the circuit participates in $(N-1)$ two-qubit controlled gates. For the gates before the Hadamard gate, the qubit serves as the control, while for those after the Hadamard gate, the qubit serves as the target.

\begin{figure}[t]
\centering
\subfloat[][The straightforward implementation]{
\includegraphics[width=.42\textwidth]{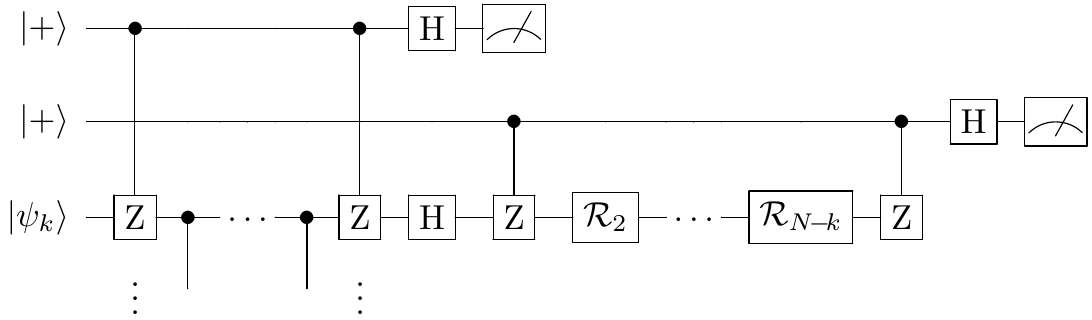}
\label{fig:sts_qft_1q}
}
\\
\subfloat[][The combined \ac{sts}]{
\includegraphics[width=.38\textwidth]{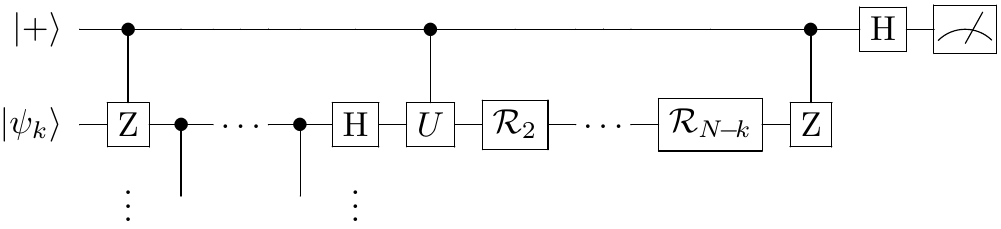}
\label{fig:sts_qft_1q_merge}
}
\caption{Implementations of the \acp{sts} on a single qubit in the \ac{qft} circuit.}
\end{figure}

We observe that for each qubit, the gates before the Hadamard gate and those after the Hadamard gate commute with the Pauli-Z operator, respectively, because all the two-qubit gates are controlled Z-rotations. Hence a straightforward implementation of the \acp{sts} is to treat these two blocks of gates separately, as shown in Fig. \ref{fig:sts_qft_1q}. However, this implementation may be excessively complex, since we would need two ancillas for every data qubit. Thus we may combine both \acp{sts} on each qubit, and arrive at the design portrayed in Fig. \ref{fig:sts_qft_1q_merge} after a slight simplification. The operator $\Sop{U}$ in Fig. \ref{fig:sts_qft_1q_merge} has the following matrix representation
\begin{equation}
\M{U} = \M{Z}\M{X} = \left[
                       \begin{array}{cc}
                         0 & -1 \\
                         1 & 0 \\
                       \end{array}
                     \right],
\end{equation}
which only differs from the Pauli-Y operator by a global phase. Note that this global phase is non-negligible in the controlled-$\Sop{U}$ operation.

\subsection{Brief Introduction to the \ac{qaoa}}\label{ssec:qaoa_intro}
The \ac{qaoa} is an algorithm aiming for approximately solving discrete optimization problems taking the following form
\begin{equation}\label{obj_qaoa_ori}
\begin{aligned}
\max_{\V{x}}& ~~F(\V{x}):=\sum_{k=1}^K f_k(\V{x}),\\
{\rm subject~to}& ~~x_i\in\{-1,1\},~\forall i=1\dotsc N,
\end{aligned}
\end{equation}
where $\V{x}=[x_1~\dotsc~x_N]^{\rm T}$, and $f_k(\V{x})$ is a $k$-th order polynomial containing only $k$-th order monomials. For example, when $N=3$, we may have $f_1(\V{x}) = 0.1x_1+0.2x_2+0.3x_3$, $f_2(\V{x})=0.4x_1x_2+0.5x_2x_3$, and $f_3(\V{x})=x_1x_2x_3$. The most common problem instances belong to the class of quadratic unconstrained binary optimization (QUBO) problems corresponding to $K=2$, which can be expressed as
\begin{equation}\label{problem_qubo}
\begin{aligned}
\max_{\V{x}}& ~~\V{x}^{\rm T}\M{A}\V{x}+\V{b}^{\rm T}\V{x},\\
{\rm subject~to}& ~~x_i\in\{-1,1\},~\forall i=1\dotsc N.
\end{aligned}
\end{equation}
By representing the vector $\V{x}$ using a quantum state $\ket{\psi}$, we could represent the objective function $F(\V{x})$ of \eqref{obj_qaoa_ori} in the following alternative form
\begin{equation}\label{obj_qaoa}
F(\ket{\psi}) = \bra{\psi}\M{H}_{\rm P}\ket{\psi},
\end{equation}
where $\M{H}_{\rm P}=\sum_{n=1}^K\M{F}_k$ is called the phase Hamiltonian encoding of the objective function, and $\M{F}_k$ is the operator obtained by replacing terms such as $x_i$ in $f_k(\V{x})$ by Pauli-Z operators $\M{Z}_i$.

In order to maximize the objective function $F(\ket{\psi})$, the \ac{qaoa} applies two Hamiltonians, namely the phase Hamiltonian and the mixing Hamiltonian, in an alternating order. Specifically, given the initial state $\ket{\psi(0)}$, the output state can be expressed as
\begin{equation}\label{qaoa_circuit}
\ket{\psi(\V{\beta},\V{\gamma})}=e^{-\imath\beta_p\M{H}_{\rm M}}e^{-\imath\gamma_p\M{H}_{\rm P}}\dotsc e^{-\imath\beta_1\M{H}_{\rm M}}e^{-\imath\gamma_1\M{H}_{\rm P}}\ket{\psi(0)},
\end{equation}
where $\V{\beta}=[\beta_1~\dotsc~\beta_p]^{\rm T}$ and $\V{\gamma}=[\gamma_1~\dotsc~\gamma_p]^{\rm T}$ are adjustable parameters controlling the search trajectory of the algorithm, and the mixing Hamiltonian $\M{H}_{\rm M}$ is given by
\begin{equation}
\M{H}_{\rm M} = \sum_{i=1}^N \M{X}_i.
\end{equation}
It has been shown that the optimal solution can be closely approximated by measuring $\ket{\psi(\V{\beta},\V{\gamma})}$ on the computational basis, when $p$ is sufficiently large and the parameters $\V{\beta}$ and $\V{\gamma}$ are chosen appropriately \cite{qaoa}.

\subsection{The \acp{sts} of the \ac{qaoa} Circuits}\label{ssec:qaoa_sts}
From \eqref{qaoa_circuit} we could observe that a typical \ac{qaoa} circuit has $p$ stages, among which the $n$-th stage is
\begin{equation}
\M{U}_n(\beta_n,\gamma_n) = e^{-\imath\beta_n\M{H}_{\rm M}}e^{-\imath\gamma_n\M{H}_{\rm P}}.
\end{equation}
Since the structure of each stage is similar, we will focus on a single stage in the following analysis. It is clear that $\M{H}_{\rm M}$ commutes with $\M{X}^{\otimes N}$ and $\M{H}_{\rm P}$ commutes with $\M{Z}^{\otimes N}$. But we could find more symmetries by decomposing the phase Hamiltonian as follows:
\begin{equation}
\begin{aligned}
\M{H}_{\rm P} = \sum_{k=1}^{\lfloor K/2 \rfloor} \M{F}_{2k} + \sum_{k=1}^{\lfloor K/2 \rfloor} \M{F}_{2k-1}  := \M{H}_{\rm P}^{(\rm even)} + \M{H}_{\rm P}^{(\rm odd)}.
\end{aligned}
\end{equation}
We note that
\begin{remark}\label{rem:odd_even}
The partial Hamiltonian $\M{H}_{\rm P}^{(\rm even)}$ corresponding to even $k$ commutes with $\M{X}^{\otimes N}$, while the part $\M{H}_{\rm P}^{(\rm odd)}$ corresponding to odd $k$ anti-commutes with $\M{X}^{\otimes N}$. Furthermore, $e^{-\imath\gamma_n\M{H}_{\rm P}^{(\rm even)}}$ also commutes with both $\M{X}^{\otimes N}$ and $\M{Z}^{\otimes N}$, since $[\M{A},\M{B}]=0$ implies $[e^{\imath\theta\M{A}},\M{B}]=0$.
\end{remark}

To see this more clearly, let us consider the QUBO case \eqref{problem_qubo}, for which we have
\begin{equation}
\begin{aligned}
\M{H}_{\rm P}^{(\rm even)} = \sum_{i=1}^N\sum_{j=1}^N a_{ij}\M{Z}_i \M{Z}_j, ~\M{H}_{\rm P}^{(\rm odd)} =\sum_{i=1}^N b_i \M{Z}_i,
\end{aligned}
\end{equation}
where $a_{ij}$ denotes the $(i,j)$-th entry of $\M{A}$ and $b_i$ denotes the $i$-th entry of $\V{b}$. Observe that the operator $\M{Z}_i\M{Z}_j$ commutes with $\M{X}^{\otimes N}$, while $\M{Z}_i$ anti-commutes with $\M{X}^{\otimes N}$.

Since the gates implementing $e^{-\imath\gamma_n\M{H}_{\rm P}}$ commute with one another, we may rearrange the order of execution of these gates, so that $e^{-\imath \gamma_n\M{H}_{\rm P}^{(\rm odd)}}$ is executed before $e^{-\imath \gamma_n\M{H}_{\rm P}^{(\rm even)}}$. This leads to the following decomposition of the $n$-th stage into three sub-stages
\begin{equation}
\begin{aligned}
\M{U}_n(\beta_n,\gamma_n) &= \M{U}_n^{(3)}\M{U}_n^{(2)}\M{U}_n^{(1)} \\
&= e^{-\imath \beta_n\M{H}_{\rm M}} e^{-\imath \gamma_n\M{H}_{\rm P}^{(\rm even)}} e^{-\imath \gamma_n\M{H}_{\rm P}^{(\rm odd)}}.
\end{aligned}
\end{equation}
This tripartite circuit has the following \acp{sts}
\begin{equation}
\M{S}_1 = \Sop{S}\{\M{Z}^{\otimes N}(0),\M{Z}^{\otimes N}(2)\},~\M{S}_2 = \Sop{S}\{\M{X}^{\otimes N}(1),\M{X}^{\otimes N}(3)\}.
\end{equation}

\begin{figure}[t]
\centering
\subfloat[][Even $N$]{
\includegraphics[width=.43\textwidth]{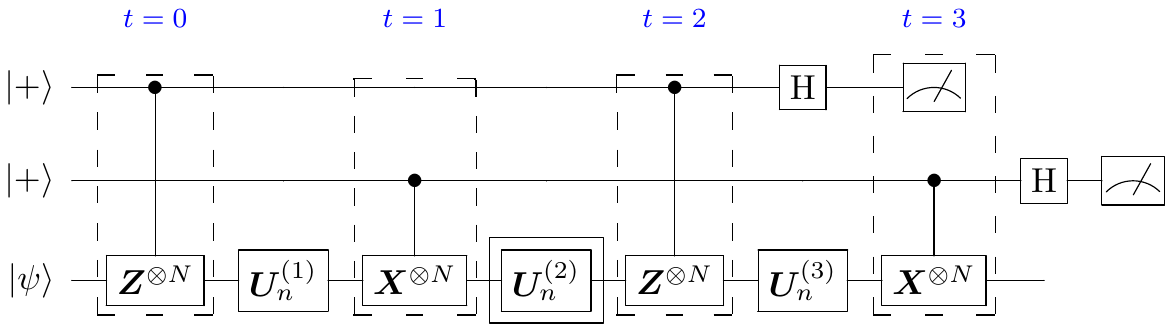}
\label{fig:sts_qaoa_even}
}
\\
\subfloat[][Odd $N$]{
\includegraphics[width=.43\textwidth]{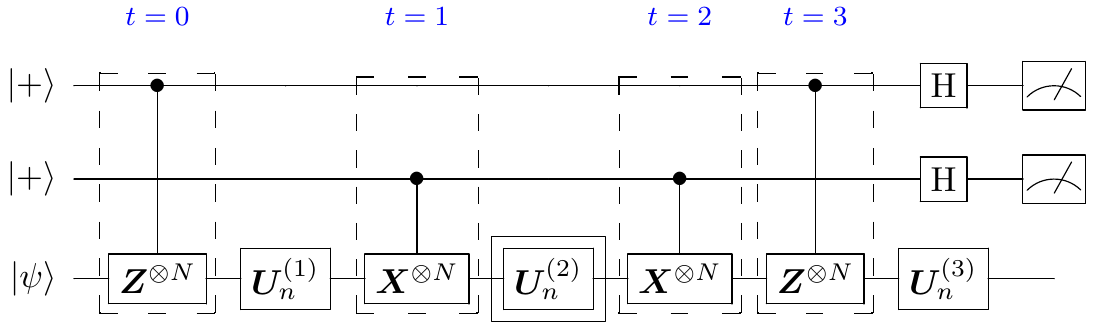}
\label{fig:sts_qaoa_odd}
}
\caption{Circuits implementing a single stage of the \ac{qaoa}, protected by two \ac{sts} checks. Especially, the gates comprising $\M{U}_n^{(2)}$ (marked by double solid lines) are protected from any single-qubit error by the \ac{sts} checks.}
\vspace{0mm}
\end{figure}

A noteworthy fact is that $\M{S}_1$ and $\M{S}_2$ are not simultaneously observable when $N$ is odd. Therefore, we arrive at different circuit implementations for even $N$ and odd $N$, as shown in Fig. \ref{fig:sts_qaoa_even} and \ref{fig:sts_qaoa_odd}, respectively. The ancillas can be re-initialized and reused in the subsequent stages. Specifically, the \acp{sts} measured in the odd $N$ scenario are
\begin{equation}
\begin{aligned}
\M{S}_1^{(\rm odd)} &= \Sop{S}\{\M{Z}^{\otimes N}(0),\M{Z}^{\otimes N}(3)\},\\
\M{S}_2^{(\rm odd)} &= \Sop{S}\{\M{X}^{\otimes N}(1),\M{X}^{\otimes N}(2)\}.
\end{aligned}
\end{equation}
Note that the action scope of $\M{S}_2^{(\rm odd)}$ lies between $\M{Z}^{\otimes N}(0)$ and $\M{Z}^{\otimes N}(3)$. Being an \ac{sts}, the insertion of $\M{S}_2^{(\rm odd)}$ does not change the partial circuit $\M{U}_n^{(2)}$. This implies that $\M{S}_1^{(\rm odd)}$ is still an \ac{sts} when $\M{S}_2^{(\rm odd)}$ is applied, and hence these two \acp{sts} are simultaneously observable.
 The main difference between the two implementations is that the third sub-stage $\M{U}_n^{(3)}$ is not protected when $N$ is odd, and thus the circuits having odd $N$ and those having even $N$ are not equally protected. Fortunately, the third sub-stage only consists of single-qubit gates that are typically less noisy in practice. Also note that the second sub-stage $\M{U}_n^{(2)} = e^{-\imath\gamma_n\M{H}_{\rm P}^{(\rm even)}}$ commutes with both $\M{X}^{\otimes N}$ and $\M{Z}^{\otimes N}$, hence we could detect any single-qubit error that occurs within this sub-stage.

\section{Numerical Results}\label{sec:numerical}
In this section, we characterize the performance of the \ac{sts} method using numerical examples. When evaluating the computational accuracy, we use the purity\footnote{Instead of evaluating directly the error of certain computational tasks, we use the purity because it does not depend on the specific observable, and hence may reflect the performance of the error mitigation techniques more clearly.} of the output state of the data register as the performance metric, defined by $\tr{\rho_{\rm data}^2}$, where $\rho_{\rm data}$ is the output state of the data register.

\subsection{Consecutive Single-Qubit Gates}
We first contrast the \ac{sts} method to the quantum switch based method described in Section \ref{sec:quantum_switch}, using the low-complexity example of single-qubit circuits. Specifically, we consider consecutive X-rotation gates applied to a single qubit. Since the gates are diagonal under the X-basis, we do not expect that any of the two methods would detect X-errors. In light of this, we assume that each X-rotation gate is associated with a Pauli-Z (dephasing) channel having the error probability of $\epsilon_1=0.001$. The two-qubit gates applied in both error mitigation methods are also assumed to be contaminated by Pauli-Z errors at an error probability of $\epsilon_2$. We will consider different values of $\epsilon_2$ in the following discussion.

\begin{figure}[t]
\centering
\subfloat[][Two consecutive gates]{
\includegraphics[width=.43\textwidth]{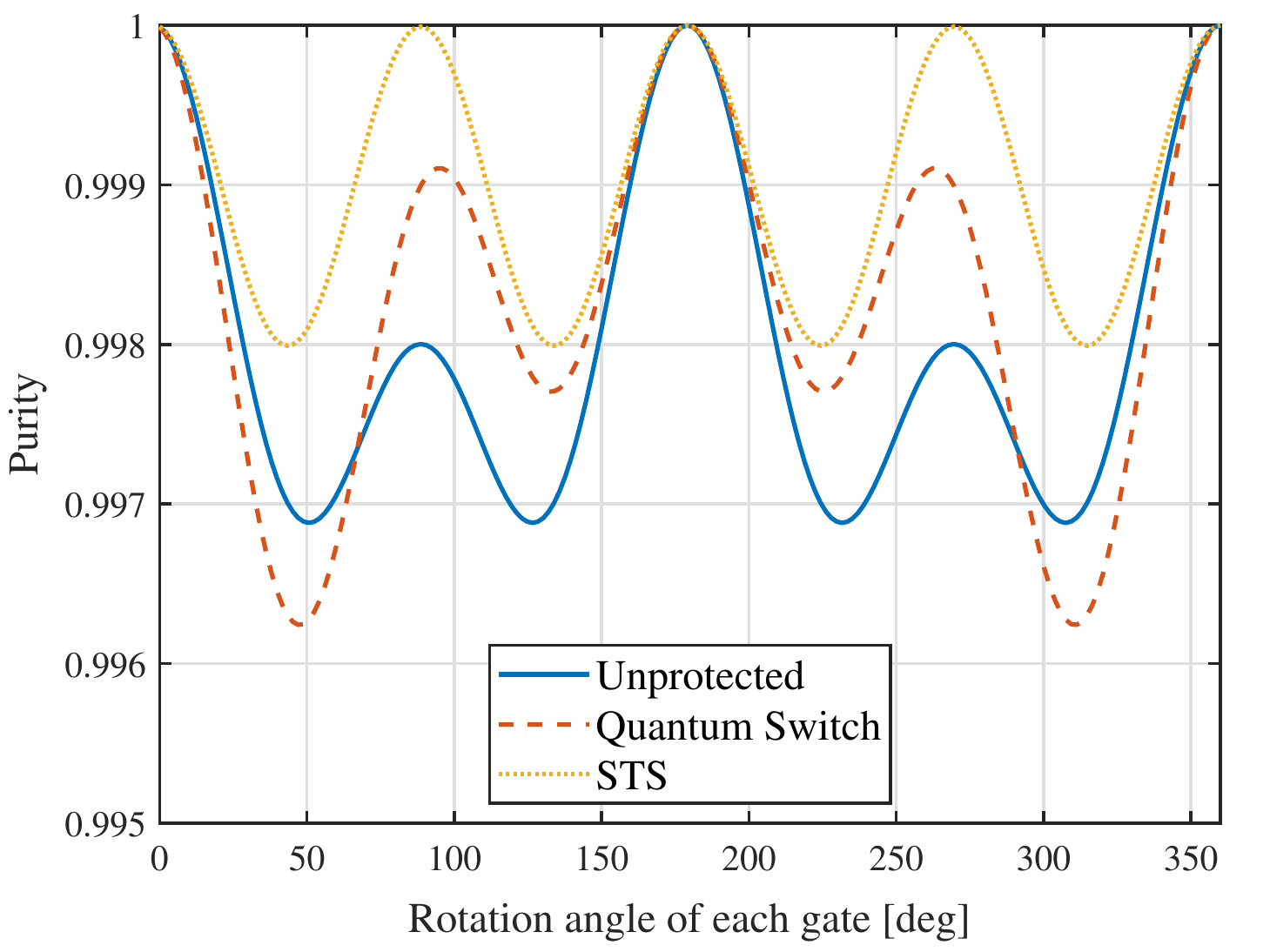}
\label{fig:repeat_gate_2_2}
}
\\
\subfloat[][Ten consecutive gates]{
\includegraphics[width=.43\textwidth]{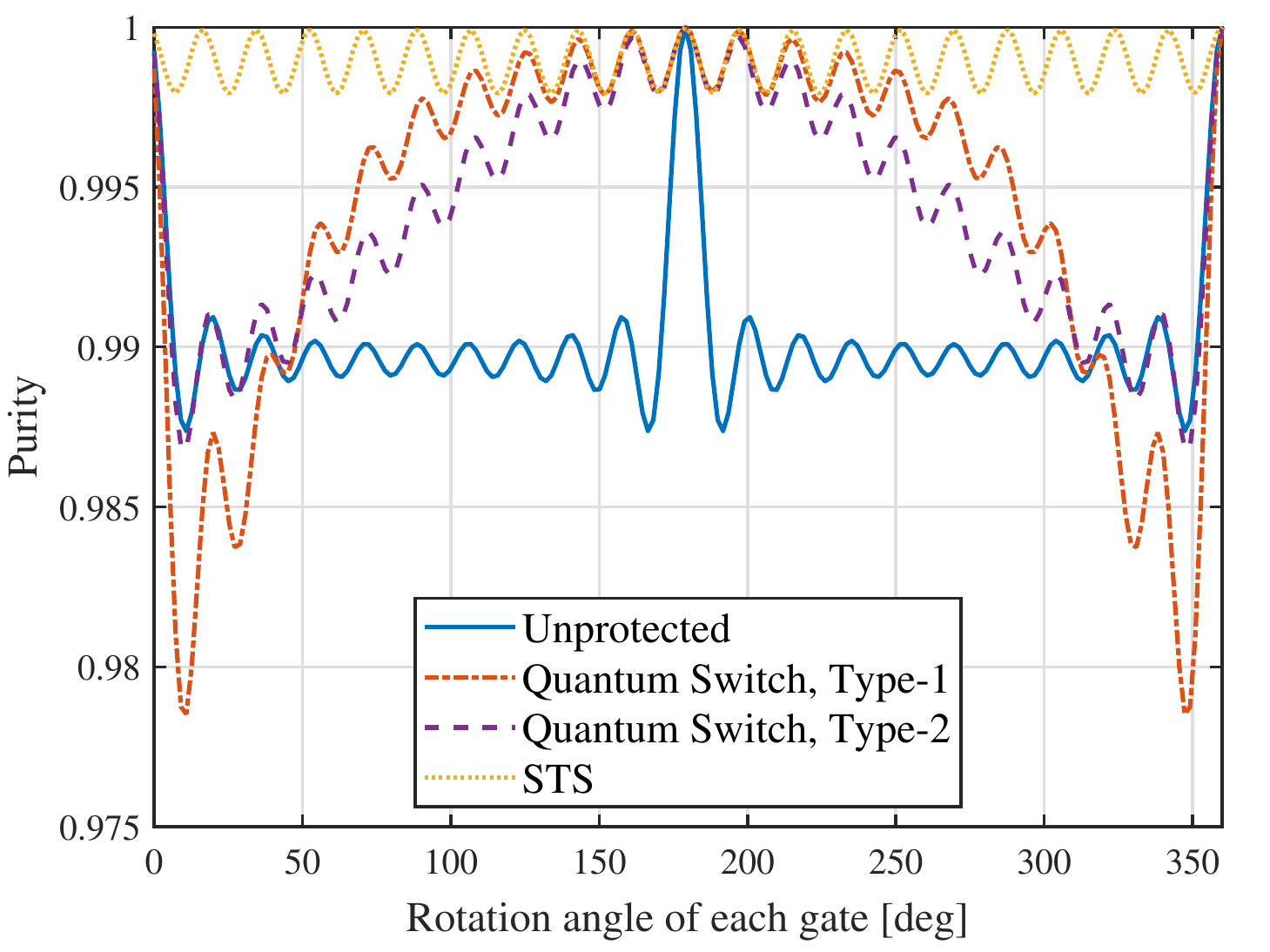}
\label{fig:repeat_gate_2_10}
}
\caption{The output purities of different implementations of consecutive X-rotation gates vs. the rotation angle of each gate, where $\epsilon_2/\epsilon_1=2$.}
\vspace{0mm}
\end{figure}

Let us first consider the case of $\epsilon_2/\epsilon_1=2$. This is an idealistic case for quantum switches, since the controlled rotation gates (e.g. the gate $A$ in Fig. \ref{fig:quantum_switch}) inflict an error on the data register at the same probability as that of the uncontrolled gates (e.g. the gate $B$ in Fig. \ref{fig:quantum_switch}). However, this is typically not the case for practical devices, for which $\epsilon_2/\epsilon_1$ is around $10$. We portray the simulation results in Fig. \ref{fig:repeat_gate_2_2} where we have $N_{\rm G}=2$ consecutive X-rotation gates, while in Fig. \ref{fig:repeat_gate_2_10} we have $N_{\rm G}=10$. As we have discussed in Section \ref{sec:qs_practical}, there are multiple possible implementations of the quantum switch based method, when $N_{\rm G}>2$. In Fig. \ref{fig:repeat_gate_2_10}, ``quantum switch, type-1'' refers to the implementation shown in Fig. \ref{fig:quantum_switch_multiple_gates}, while ``quantum switch, type-2'' refers to that shown in Fig. \ref{fig:quantum_switch_multiple_gates2}.

Observe from Fig. \ref{fig:repeat_gate_2_2} that the output purity of both the quantum switch and of the \ac{sts} depends on the rotation angle of each X-rotation gate. To elaborate, the rotation angle has an impact on the commutativity with the Z-error, which in turn determines the error mitigation performance. Observe from Fig. \ref{fig:repeat_gate_2_10} that, compared to the unprotected circuits, the accuracy improvement of both methods becomes more significant when $N_{\rm G}$ is larger, since the additional error introduced by the methods themselves becomes less severe than that of the consecutive X-rotations. An interesting phenomenon is that the quantum switch based method performs better for larger rotation angles. This may be interpreted as a penalty of treating the X-rotation gate itself as the reference of symmetry verification, instead of using a universal reference (e.g. Pauli-X operators in the \ac{sts} method).

\begin{figure}[t]
\centering
\subfloat[][Two consecutive gates]{
\includegraphics[width=.43\textwidth]{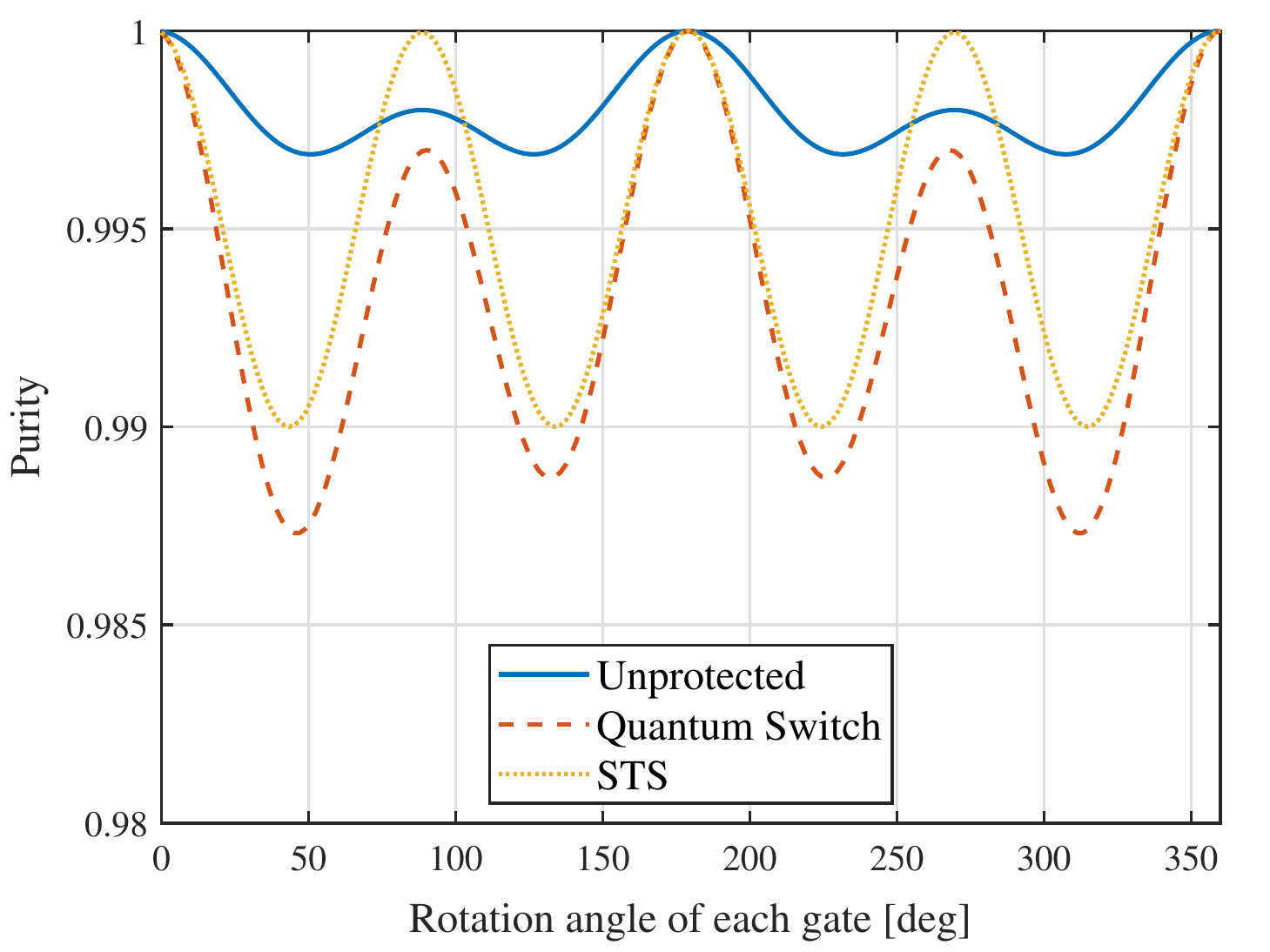}
\label{fig:repeat_gate_10_2}
}
\\
\subfloat[][Ten consecutive gates]{
\includegraphics[width=.43\textwidth]{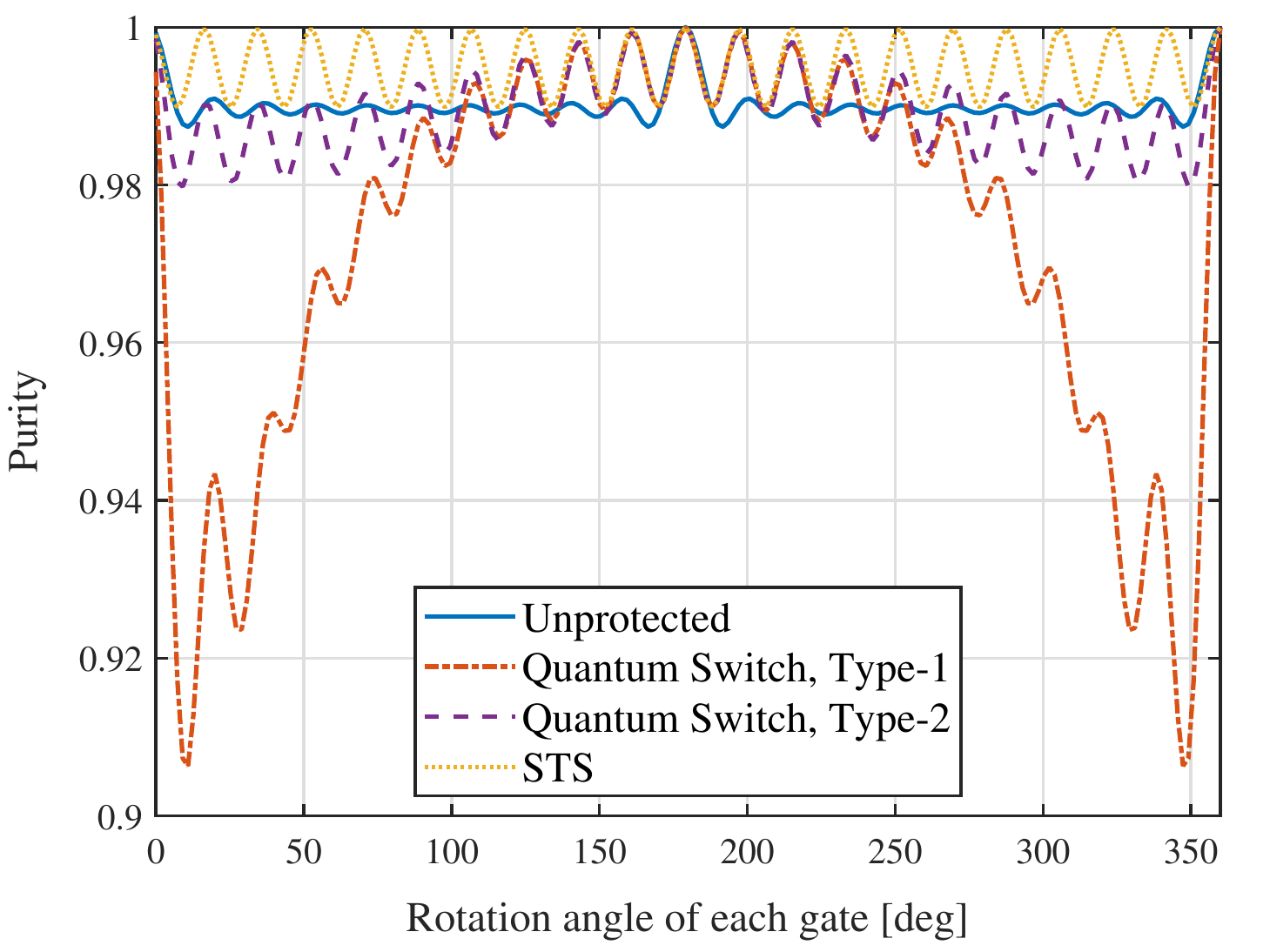}
\label{fig:repeat_gate_10_10}
}
\caption{The output purities of different implementations of consecutive X-rotation gates vs. the rotation angle of each gate, where $\epsilon_2/\epsilon_1=10$.}
\vspace{0mm}
\end{figure}

The results are portrayed for the more practical case of $\epsilon_2/\epsilon_1=10$ in Fig. \ref{fig:repeat_gate_10_2} and \ref{fig:repeat_gate_10_10}. We see that the quantum switch based method is only beneficial for a limited range of rotation angles in the $N_{\rm G}=10$ case, while \ac{sts} is beneficial across a wider range. Note that the \ac{sts} technique may be generalized to more complex circuits. Hence may expect that \acp{sts} are potentially beneficial for a large range of practical circuits, while quantum switches might only be useful for certain special circuits. However, it is noteworthy that using \acp{sts} requires the knowledge of the specific type of symmetry, while quantum switches are applicable as long as we know that certain gates commute with each other.

\begin{table}[h]
\centering
\caption{The output purities of different methods under bit-flip channels with error probability $0.001$.}
\begin{tabular}{|c|c|c|c|c|}
  \hline
   & \tabincell{c}{2 gates, \\$\frac{\epsilon_2}{\epsilon_1}=2$} & \tabincell{c}{2 gates, \\ $\frac{\epsilon_2}{\epsilon_1}=10$} & \tabincell{c}{10 gates, \\ $\frac{\epsilon_2}{\epsilon_1}=2$} & \tabincell{c}{10 gates, \\ $\frac{\epsilon_2}{\epsilon_1}=10$} \\ \hline
  Unprotected & 0.9960 & 0.9960 & 0.9804 & 0.9804 \\ \hline
  QS (original) & 0.9940 & 0.9784 & n/a & n/a \\ \hline
  QS type-1 & n/a & n/a & 0.9786 & 0.8468 \\ \hline
  QS type-2 & n/a & n/a & 0.9746 & 0.9608 \\ \hline
  STS & 0.9900 & 0.9670 & 0.9634 & 0.9523 \\
  \hline
\end{tabular}
\label{tbl:x_error}
\vspace{-3mm}
\end{table}

Finally, let us investigate the performance of the methods considered under the X-error model. Specifically, we consider bit-flip channels with error probability $0.001$. Since all the methods considered are incapable of detecting X-errors, the output purities are constant with respect to the rotation angles of each gate. Hence we collected the output purities of different circuits in Table I. Observe that the unprotected circuit always has the highest output purities, since all other methods would apply more gates. In particular, for the circuit constituted by two consecutive gates, the STS has lower purity than that of the quantum switch, since the former has a larger gate count (by 2). For the circuit constituted by ten consecutive gates, we see that the type-1 quantum switch has the lowest purity, since its circuit implementation (see Fig.~\ref{fig:quantum_switch_multiple_gates}) is far more complicated than other methods.

\subsection{\ac{qft} Circuits}\label{ssec:numerical_qft}
In this subsection, we evaluate the error mitigation performance of \acp{sts} when applied to $N$-qubit \ac{qft} circuits.

\begin{figure}[t]
\centering
\subfloat[][The purities]{
\includegraphics[width=.43\textwidth]{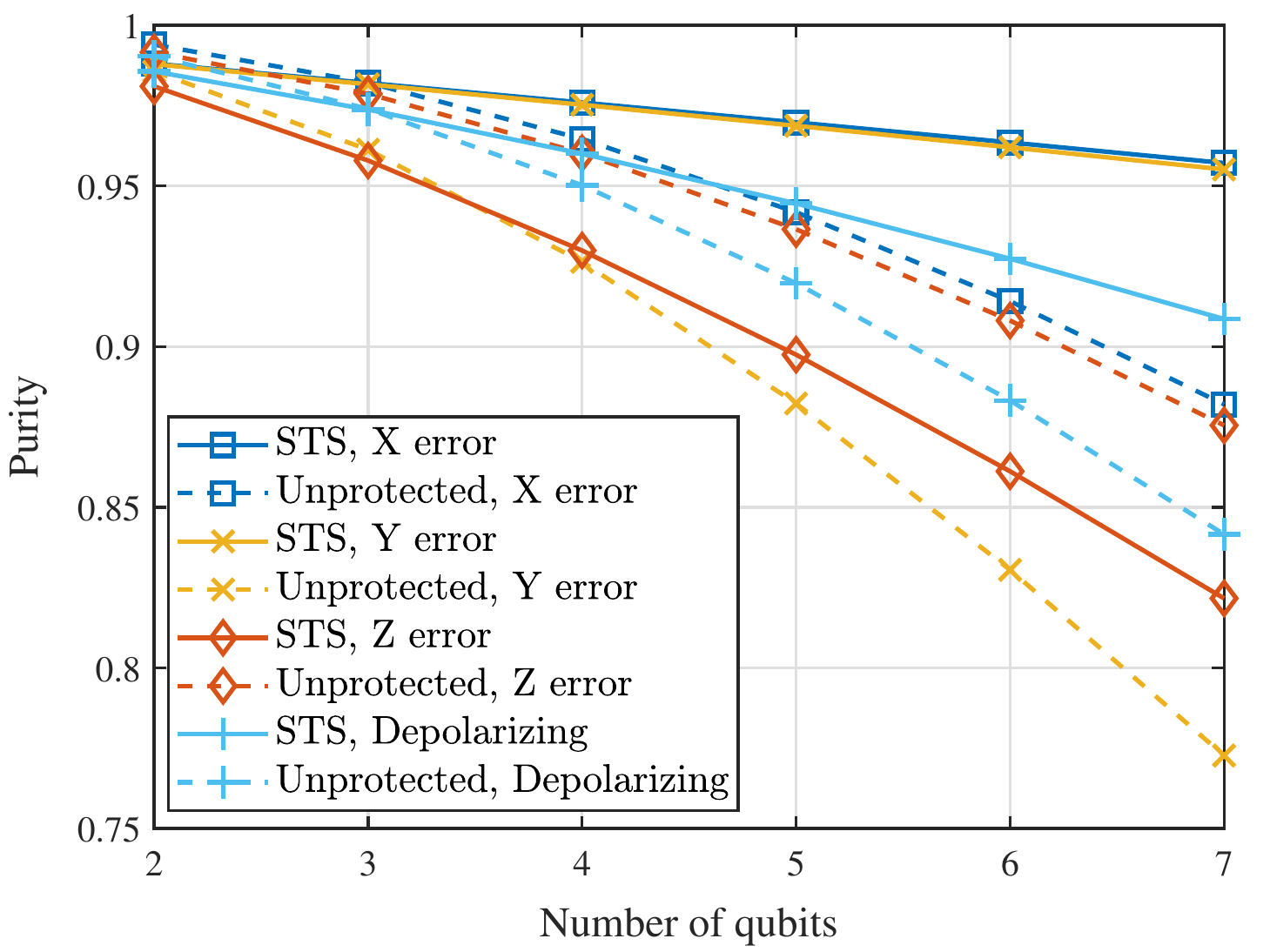}
\label{fig:qft_nq}
}
\\
\subfloat[][The sampling overhead factors]{
\includegraphics[width=.43\textwidth]{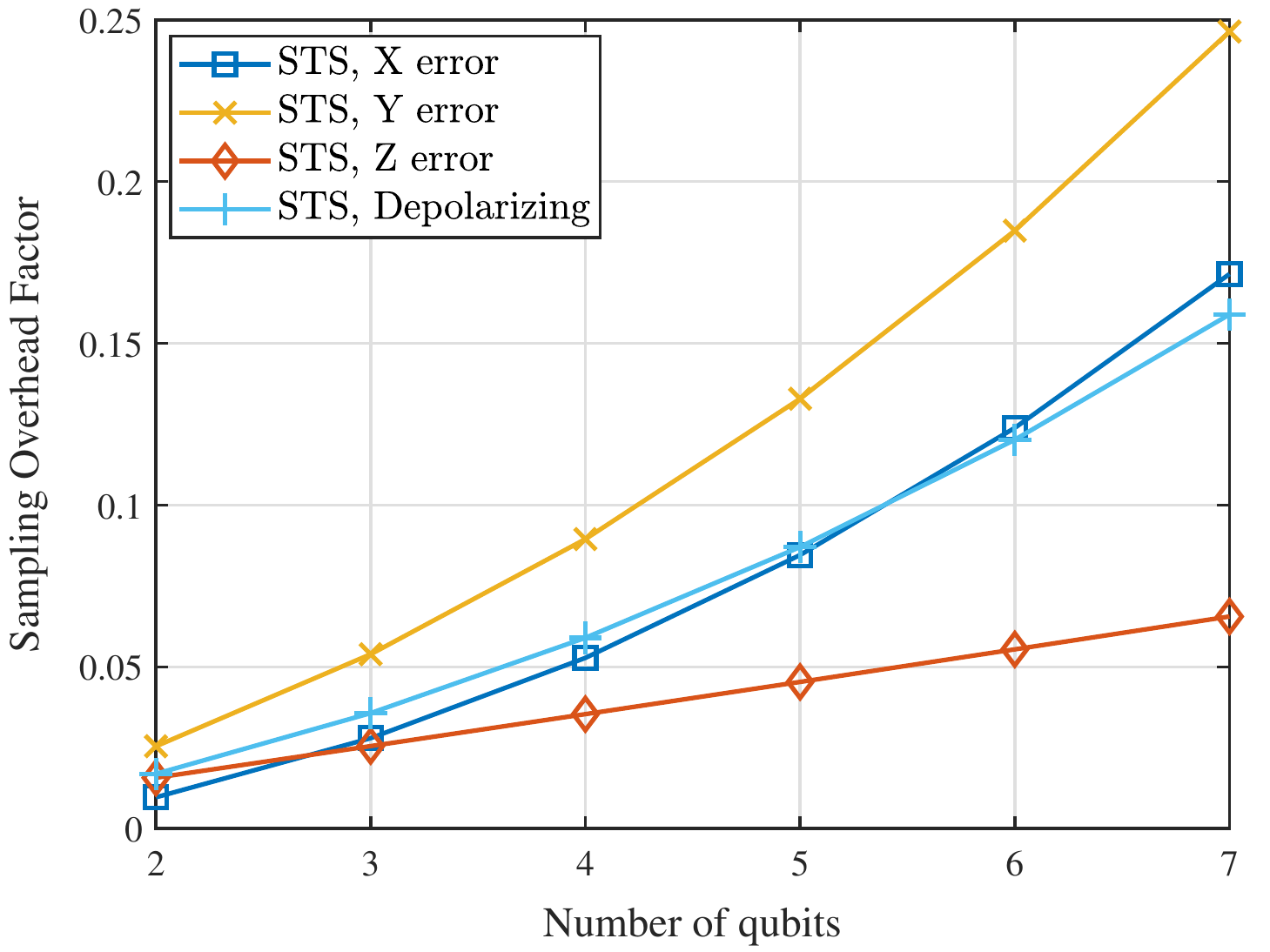}
\label{fig:qft_nq_sof}
}
\caption{The output purities and the sampling overhead factors of \ac{qft} circuits under different channel models, as functions of the number of qubits. The gate error rate is $0.003$ for two-qubit gates, and $0.0003$ for single-qubit gates.}
\vspace{0mm}
\end{figure}

Specifically, we consider the combined \ac{sts} shown in Fig. \ref{fig:sts_qft_1q_merge}. The output purities under various channel models are shown in Fig. \ref{fig:qft_nq}. Observe that \acp{sts} are more beneficial under Y-error as well as X-error channels, and they are even detrimental for Z-error channels. This is as expected, since the \acp{sts} of \ac{qft} circuits commute with Z-errors. As for the sampling overhead, it is seen from Fig. \ref{fig:qft_nq_sof} that the sampling overhead factor increases with the error detection probability, as may be inferred from its definition \eqref{def_sof}.

\begin{figure}[t]
\centering
\subfloat[][The purities]{
\includegraphics[width=.43\textwidth]{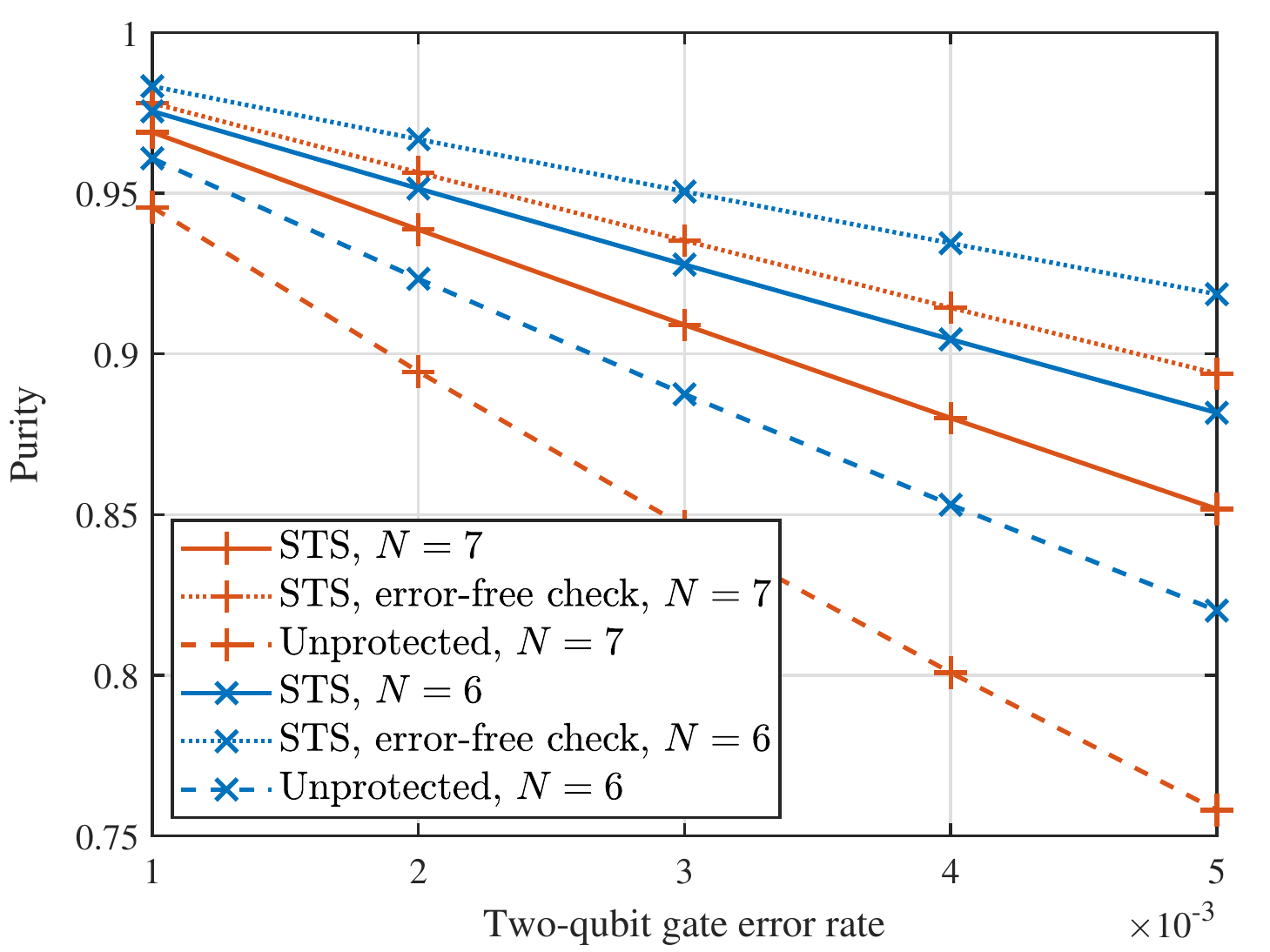}
\label{fig:qft_errrate}
}
\\
\subfloat[][The sampling overhead factors]{
\includegraphics[width=.43\textwidth]{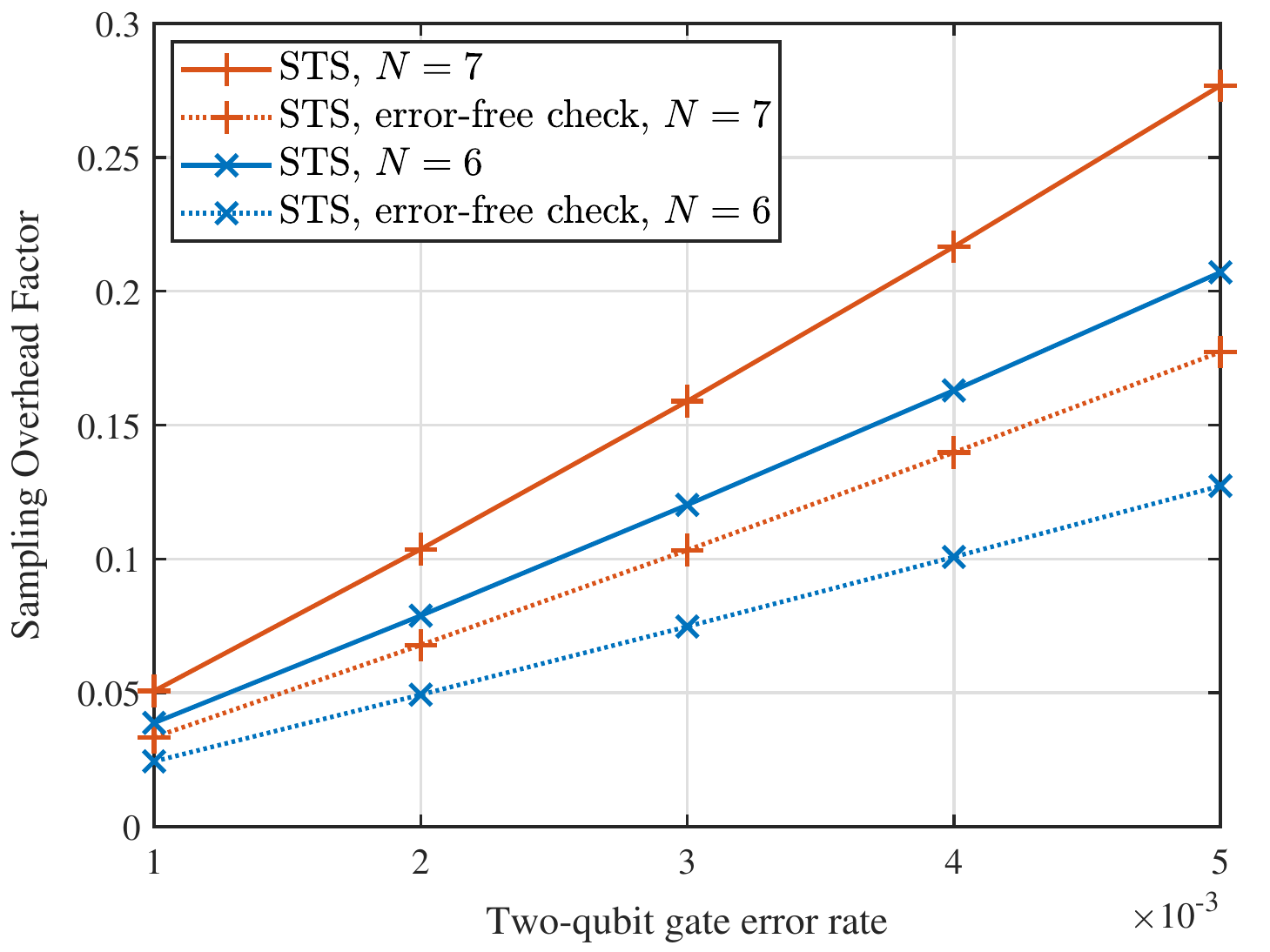}
\label{fig:qft_errrate_sof}
}
\caption{The output purities and the sampling overhead factors of \ac{qft} circuits under depolarizing channels, as functions of the error rate of two-qubit gates. The error rate of single-qubit gates is $1/10$ that of two-qubit gates.}
\vspace{0mm}
\end{figure}

The output purity versus the gate error rate under depolarizing channels is illustrated in Fig. \ref{fig:qft_errrate}. Here we consider the practical case of $\epsilon_2/\epsilon_1=10$, where $\epsilon_1$ and $\epsilon_2$ are the error rates of single-qubit and two-qubit gates, respectively.  The curves marked by ``\ac{sts}, error-free check'' correspond to the idealistic case where the gates used for implementing \ac{sts} checks are error-free. We see that the purity decreases approximately linearly as the gate error rate increases. It is also noteworthy that the purity decreases faster for larger $N$, since the number of gates is also larger.

We conclude that, for \ac{qft} circuits, the \ac{sts} method is particularly beneficial for asymmetric channels, for example, when the rate of X-errors  is $10$ times that of Z-errors. Note that the specific type of the error does not matter as long as the channel is asymmetric, because we may apply a global rotation to the entire circuit for ensuring that the dominant type of errors does not commute with the gates.

\subsection{\ac{qaoa} Circuits}
Finally, let us evaluate the performance of \acp{sts} applied to \ac{qaoa} circuits discussed in Section \ref{ssec:qaoa_intro} and \ref{ssec:qaoa_sts}. We first consider single-stage \ac{qaoa} circuits, denoted as \ac{qaoa}$_1$ circuits. For the simulations in this subsection, we use the following phase Hamiltonian
\begin{equation}
\M{H}_{\rm P}= \sum_{i=1}^N\sum_{j=1}^N a_{ij}\M{Z}_i \M{Z}_j + \sum_{i=1}^N b_i \M{Z}_i,
\end{equation}
where $a_{ij}$ and $b_i$ are randomly drawn from the uniform distribution {over the interval} $(-1,1)$. The simulation results are then averaged over $1000$ random instances of the parameters. Every two-qubit gate is affected by a depolarizing channel having a depolarizing probability of $0.001$, while the single-qubit gates have $10$ times lower depolarizing probabilities.

\begin{figure}[t]
\centering
\subfloat[][The purities]{
\includegraphics[width=.43\textwidth]{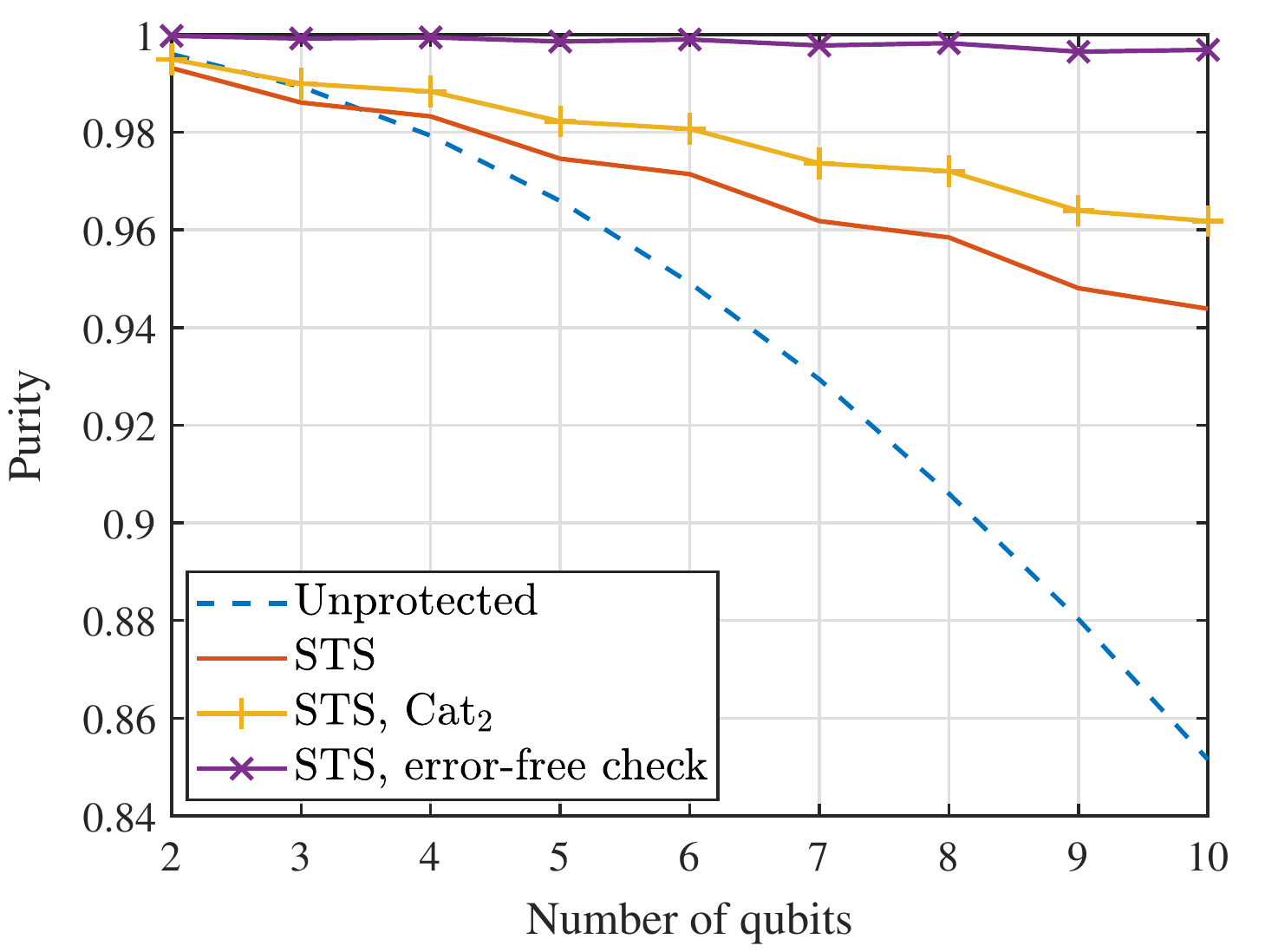}
\label{fig:qaoa1}
}
\\
\subfloat[][The sampling overhead factors]{
\includegraphics[width=.43\textwidth]{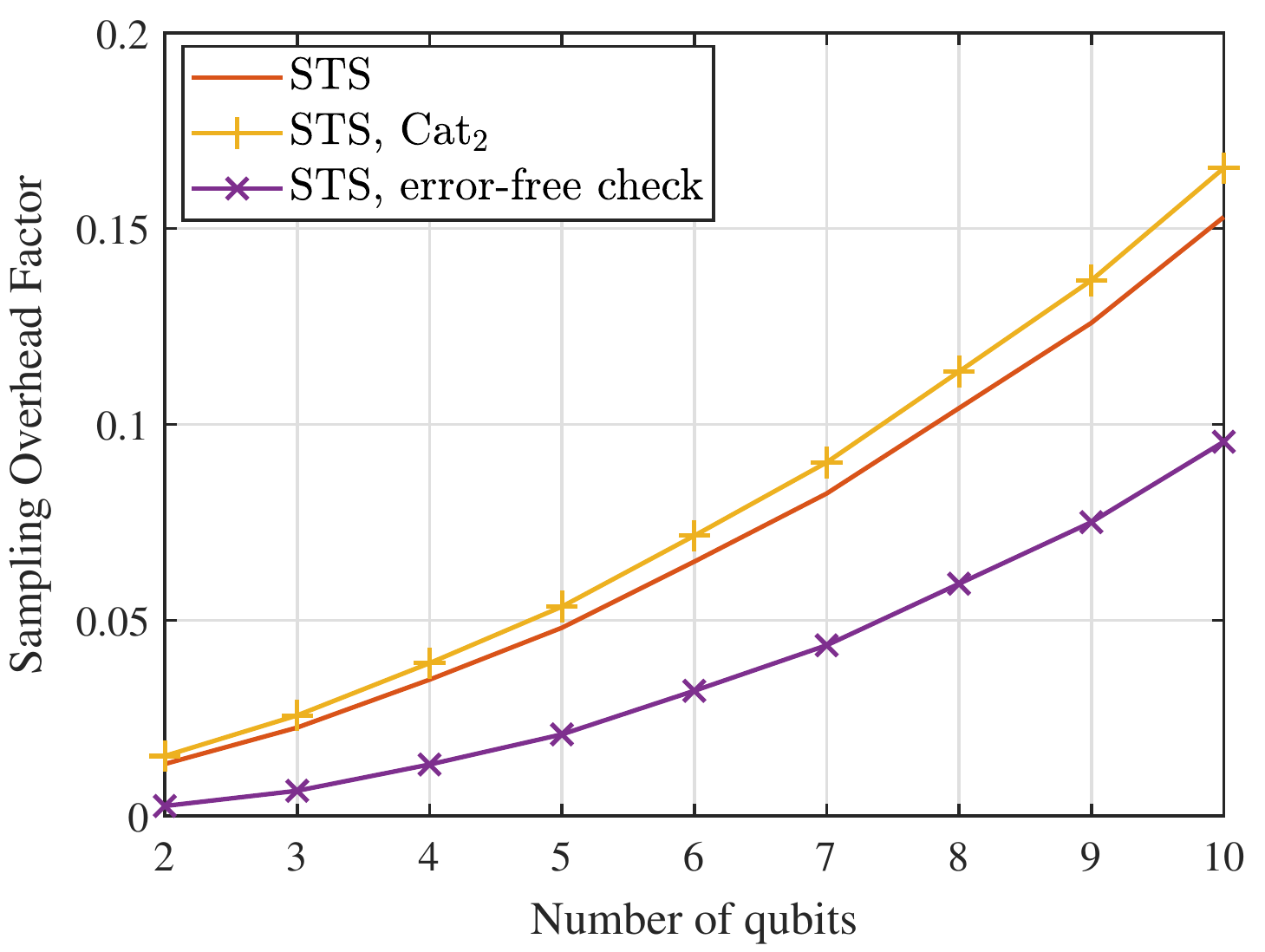}
\label{fig:qaoa1_sof}
}
\caption{The output purities and the sampling overhead factors of different implementations of the \ac{qaoa}$_1$ circuit, as functions of the number of qubits.}
\vspace{0mm}
\end{figure}

\begin{figure*}[t]
\centering
\includegraphics[width=.8\textwidth]{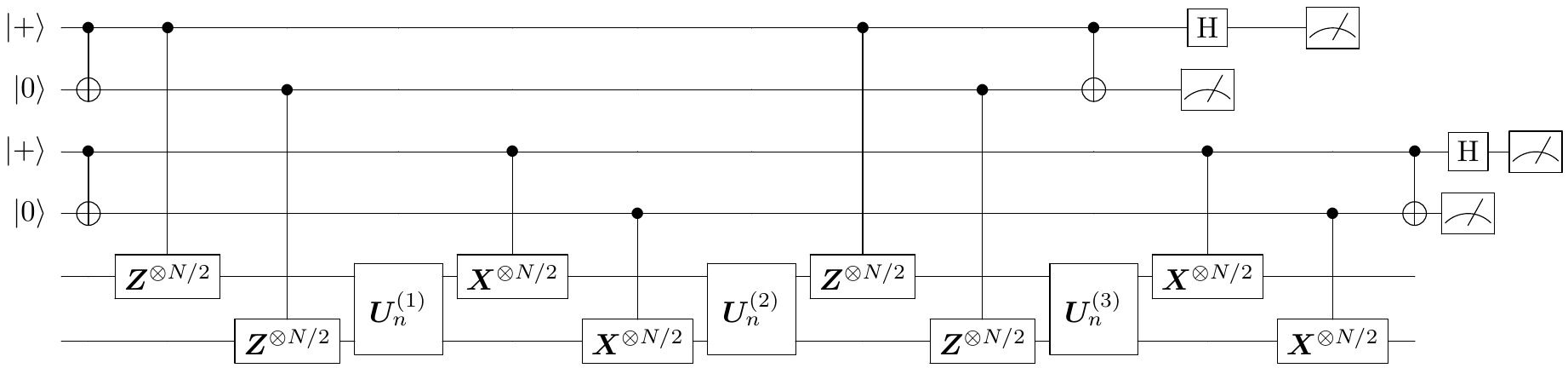}
\vspace{0mm}
\caption{The implementation of an \ac{sts}-protected \ac{qaoa}$_1$ circuit, relying on two control qubits.}
\label{fig:sts_qaoa_2a}
\vspace{0mm}
\end{figure*}

The output purities and the sampling overhead factors are shown in Fig. \ref{fig:qaoa1} and \ref{fig:qaoa1_sof}, respectively. In these figures, ``\ac{sts}, Cat$_2$'' refers to the implementation of \acp{sts} relying on cat states defined on two ancillas, as portrayed in Fig. \ref{fig:split_sts}. The specific implementation of \ac{qaoa} circuits is portrayed in Fig. \ref{fig:sts_qaoa_2a}.

Observe from Fig. \ref{fig:qaoa1} that the \ac{sts} method relying on cat states defined on two ancillas outperforms its counterpart relying on a single ancilla. This corroborates with our discussion on the mitigation of error proliferation in Section \ref{ssec:accuracy_vs_overhead}, and demonstrates the trade-off between accuracy and qubit overhead. The sampling overhead factors shown in Fig. \ref{fig:qaoa1_sof} are on the order of the corresponding error detection probability, similar to our previous discussion on \ac{qft} circuits in Section \ref{ssec:numerical_qft}.

\begin{figure}[t]
\centering
\subfloat[][The purities]{
\includegraphics[width=.43\textwidth]{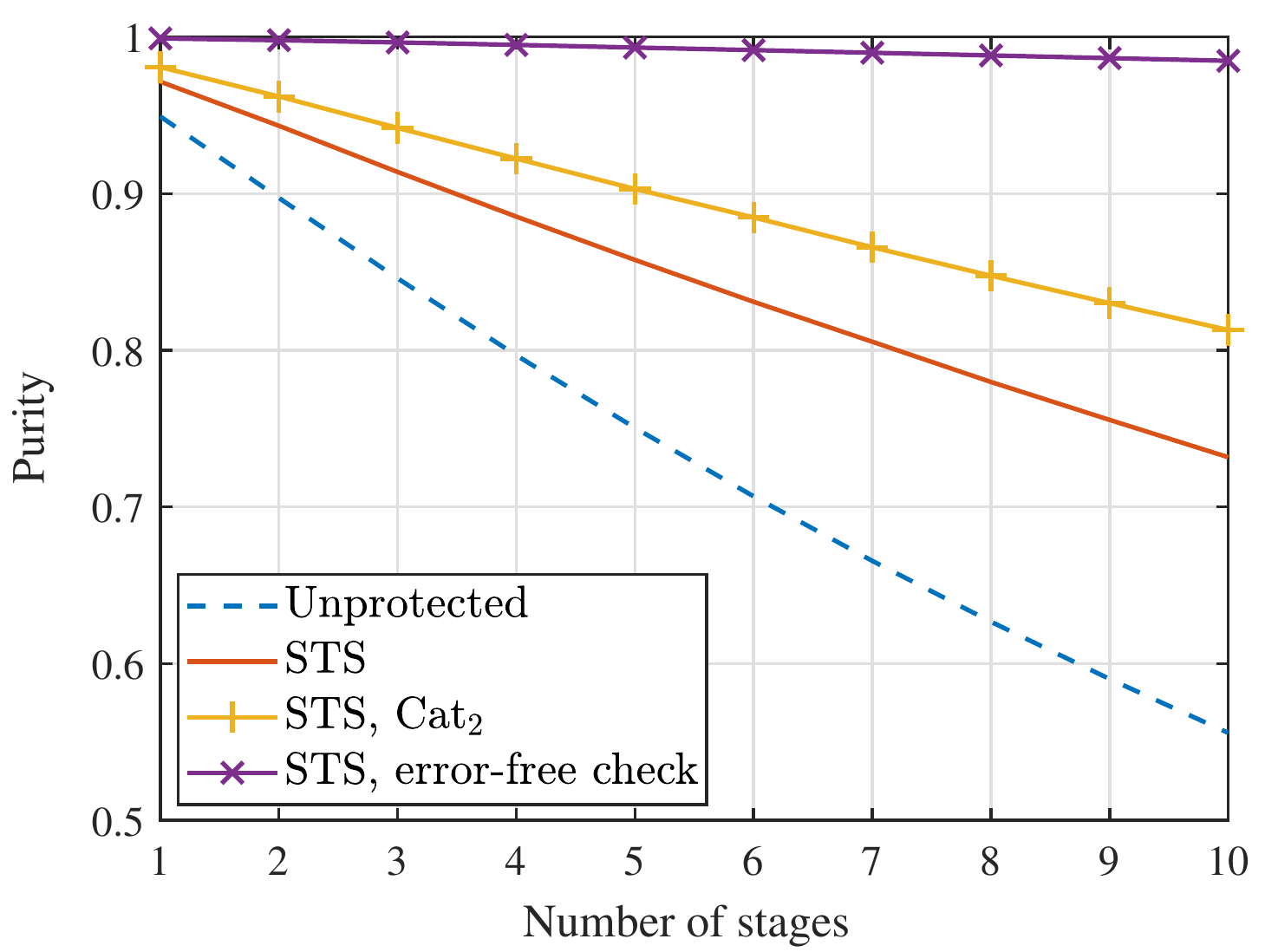}
\label{fig:qaoa_multistage}
}
\\
\subfloat[][The sampling overhead factors]{
\includegraphics[width=.43\textwidth]{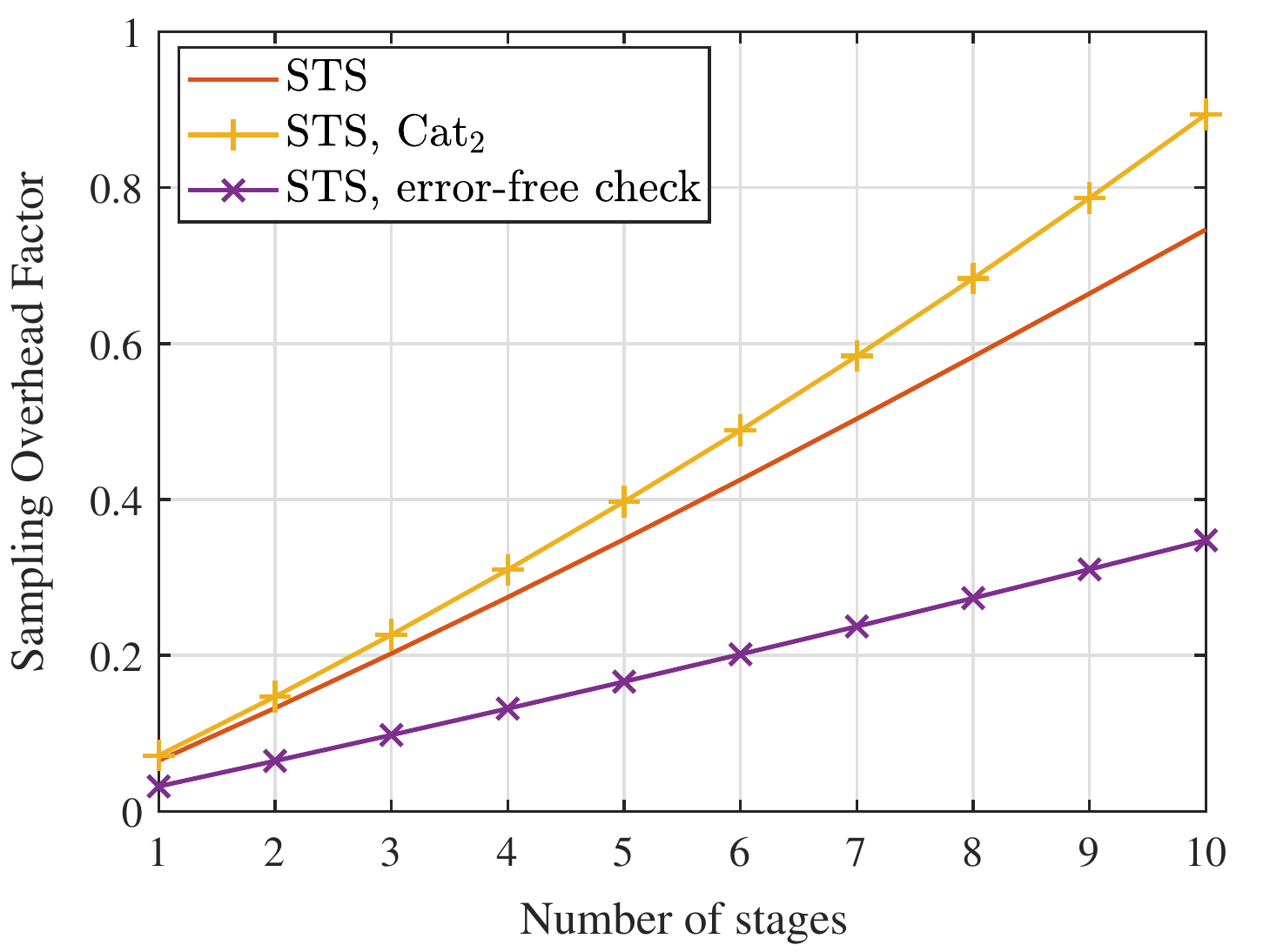}
\label{fig:qaoa1_multistage_sof}
}
\caption{The output purities and the sampling overhead factors of different implementations of multistage \ac{qaoa} circuits, as functions of the number of stages.}
\vspace{0mm}
\end{figure}

Note that the purity curves of \ac{sts} methods in Fig. \ref{fig:qaoa1} are not smooth. This is due to the fact that \ac{qaoa} circuits relying on an even number of qubits and those on an odd number of qubits are not equally protected. Indeed, as we may observe from Fig. \ref{fig:sts_qaoa_even} and \ref{fig:sts_qaoa_odd}, the final sub-stage corresponding to the mixing Hamiltonian is not protected, when the number of data qubits $N$ is odd, which is due to the simultaneous observability issue of the \acp{sts}. Consequently, the purities of \ac{qaoa} circuits having odd $N$ are lower than the expected purity, when the simultaneous observability is not an issue.

Next we consider multistage \ac{qaoa} circuits. The components of the parameter vectors $\V{\alpha}$ and $\V{\beta}$ are randomly drawn from uniform distributions on $(-\pi,\pi)$. As it can be seen from Fig. \ref{fig:qaoa_multistage}, the purity of the cat-state \ac{sts} method decreases more slowly than that of the \ac{sts} method relying on a single ancilla. Due to the complexity escalation of emulating quantum circuits on classical computers, we cannot produce the results of the \ac{sts} method relying on larger cat states defined on $N_{\rm c}>2$ ancillas. We conjecture that the purity can be further improved by using more ancillas, which is ultimately upper-bounded by the purity when the gates used for \ac{sts} checks are error-free.

\subsection{Experimental Results}\label{ssec:exp}
\begin{figure}
\centering
\subfloat[][Original circuit]{
\includegraphics[width=.46\textwidth]{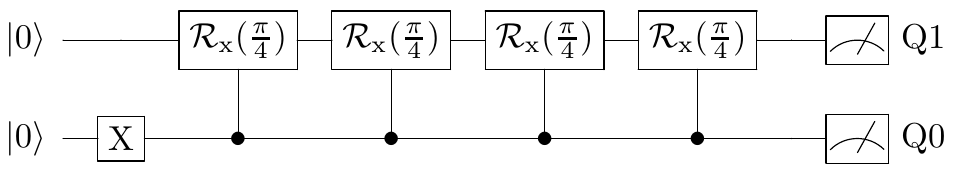}
\label{fig:exp_original}
}
\hspace{5mm}
\subfloat[][\ac{sts}-protected circuit]{
\includegraphics[width=.46\textwidth]{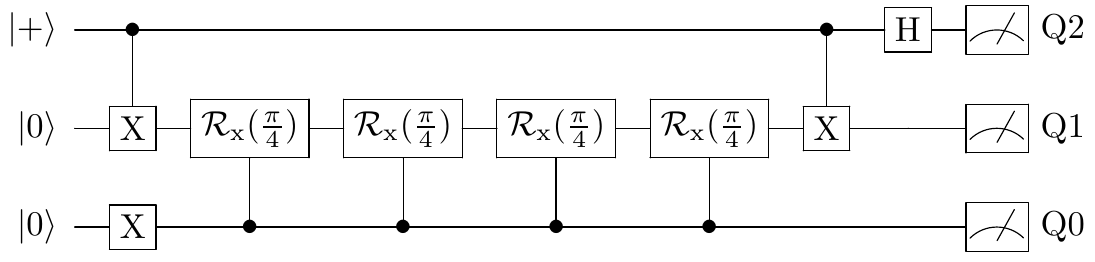}
\label{fig:exp_sts}
}
\caption{Quantum circuits used in the experiments.}
\end{figure}

To further validate and demonstrate the performance of \acp{sts}, we have conducted experiments on IBM's quantum computer IBMQ\_Lima, which is available in open access \cite{ibmq}. In particular, we consider a quantum circuit constituted by four consecutive controlled-$\mathcal{R}_{\rm x}(\pi/4)$ gates, as portrayed in Fig. \ref{fig:exp_original}. When the circuit is free of error, we expect to measure $\ket{01}$ at its output.

\begin{figure}[t]
\centering
\includegraphics[width=.24\textwidth]{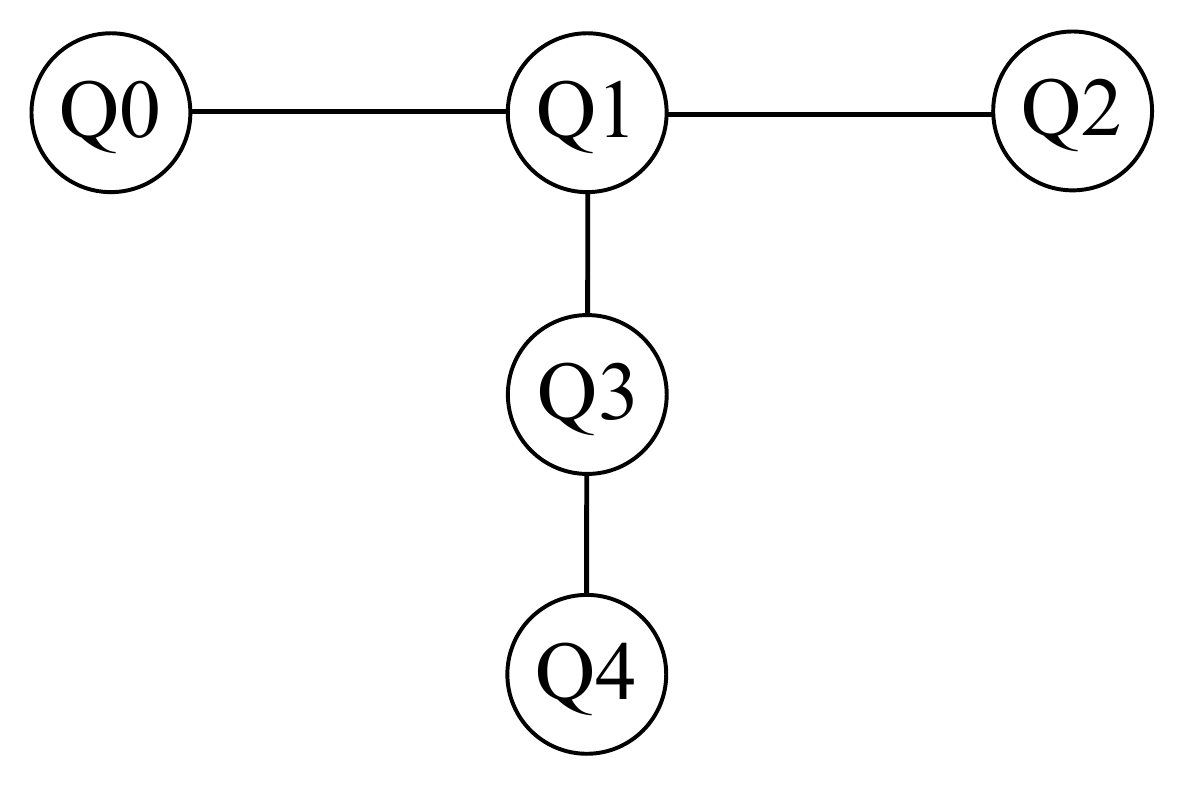}
\caption{The qubit arrangement of the quantum computer IBMQ\_Lima.}
\label{fig:ibmq_lima}
\end{figure}

This circuit clearly commutes with the operator $\M{X}\otimes \M{X}$, hence has the \ac{sts}
$$
\mathcal{S}^{(\rm st)}\{\M{X}_1(0),\M{X}_2(0),\M{X}_1(1),\M{X}_2(1)\}.
$$
Measuring this \ac{sts} poses an implicit requirement that the ancilla should be able to perform two-qubit interactions (gates) with two other qubits.  However, in IBMQ\_Lima, there is no three-qubit group in which every pair of qubits are connected, as may be observed from Fig.~\ref{fig:ibmq_lima}. In light of this, we measure the following \ac{sts}
$$
\mathcal{S}^{(\rm st)}\{\M{X}_2(0),\M{X}_2(1)\},
$$
using the circuit shown in Fig.~\ref{fig:exp_sts}, which only protects the second qubit (Q1). It is thus expected that although the error rate on Q1 may be reduced, the error rate on Q0 would even be higher after the \ac{sts} measurement, since the total number of gates is increased.

\begin{figure}[t]
\centering
\includegraphics[width=.46\textwidth]{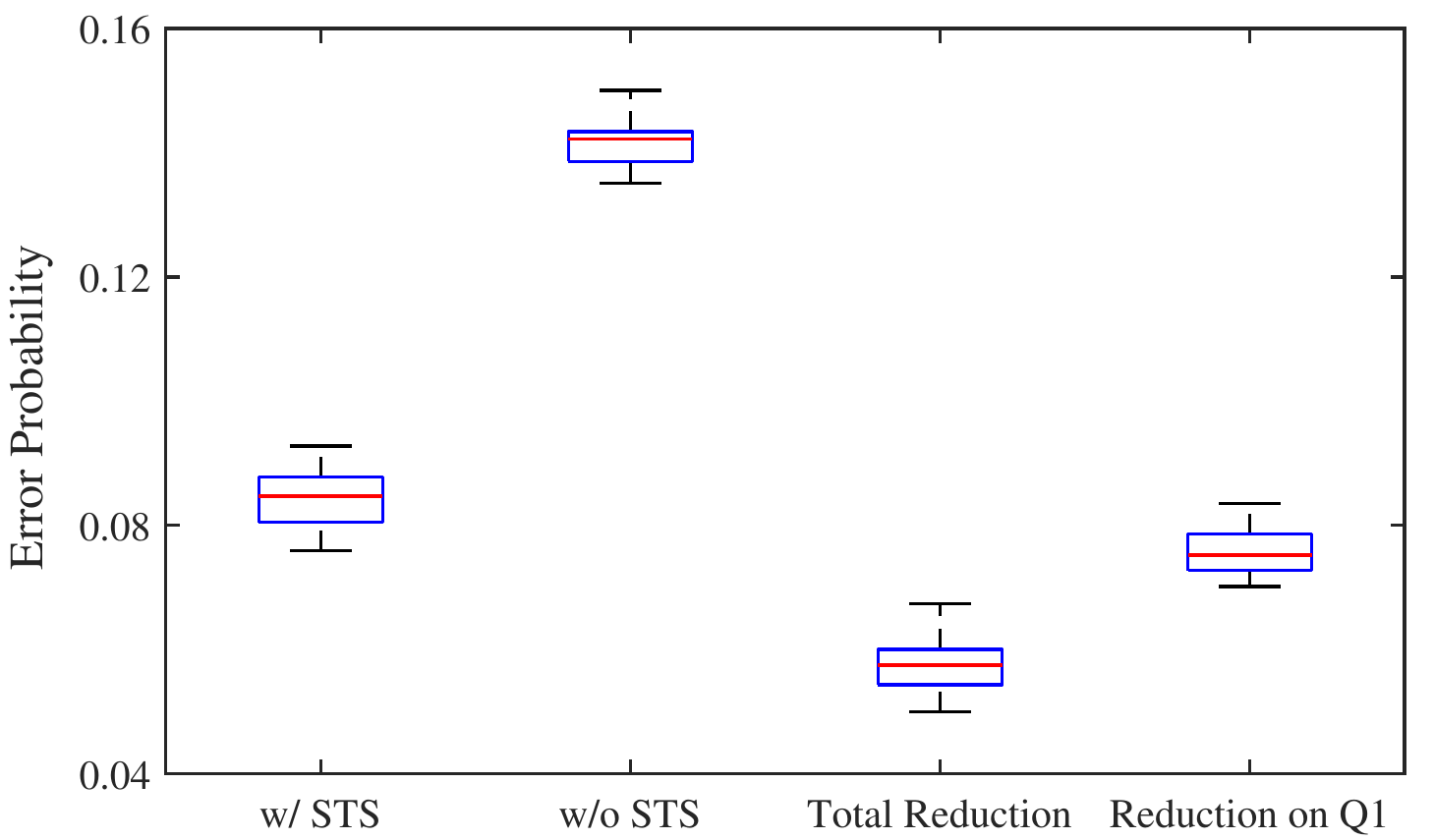}
\caption{The output error probability observed in the experiments conducted on IBMQ\_Lima.}
\label{fig:sts_xx4_ibmq_lima}
\end{figure}

We repeated the experiment $20$ times. In each experiment, we activated both the original circuit shown in Fig.~\ref{fig:exp_original} and the \ac{sts}-protected circuit shown in Fig.~\ref{fig:exp_sts}, each for $N_{\rm s}=20000$ times. We compute the error probability of the original circuit according to
\begin{equation}
p_{\rm e}^{\rm ori}=\frac{N_{\rm s}-N_{01}}{N_{\rm s}},
\end{equation}
where $N_{01}$ denotes the number of circuit activations that output $\ket{01}$. The error probability of the \ac{sts}-protected circuit is computed as
\begin{equation}
p_{\rm e}^{\rm sts} = \frac{N_0^{\rm anc}-N_{001}}{N_0^{\rm anc}},
\end{equation}
where $N_0^{\rm anc}$ denotes the number of circuit activations in which the ancilla outputs $\ket{0}$. By contrast, $N_{001}$ represents the number of circuit activations, when the entire output state is $\ket{001}$.

As it may be observed from Fig.~\ref{fig:sts_xx4_ibmq_lima}, the average error probability (over $20$ experiments) without the protection of \ac{sts} is around $14.2$\%, while the average error probability of the \ac{sts}-protected circuit is around $8.5$\%, with a total error reduction of $5.7$\%. We also note that the error reduction on Q1 is around $7.5$\%, which is significantly higher than the total reduction, due to the increased number of gates applied on Q1 in the \ac{sts}-protected circuit.

\section{Conclusions}\label{sec:conclusions}
In this treatise, we have proposed a general framework for circuit-oriented symmetry verification. Specifically, the quantum switch based method can be directly applied, when certain gates are known to commute with each other. For the case where the circuit has known symmetries, we propose the method of \ac{sts}, generalizing the concept of conventional stabilizers used for state-oriented symmetry verifications. This method is capable of verifying the symmetries without the knowledge of the current quantum state. Another major difference between \acp{sts} and their conventional counterparts is that they are not necessarily simultaneously observable, and hence sometimes a rearrangement of the circuit is required to perform multiple \ac{sts} checks. We have also discussed the accuracy vs. overhead trade-off of \acp{sts}, and provided quantum circuit designs that strike flexible trade-offs. Finally, we have demonstrated the performance of the proposed methods using numerical examples concerning practical quantum algorithms, including the \ac{qft} and the \ac{qaoa}. A possible future research direction is to find more practical algorithms for which \acp{sts} is beneficial.

\section*{Acknowledgments}
The authors acknowledge the use of the IRIDIS High Performance Computing Facility and associated support services at the University of Southampton, and the quantum simulation toolkit QuESTlink \cite{questlink}, in the completion of this work.

\appendices
\section{Proof of Proposition \ref{prop:qs}}\label{sec:proof_qs}
\begin{IEEEproof}
When $\omega=\ket{+}\!\bra{+}$, we have
\begin{equation}
\begin{aligned}
&\M{C}_{ij}(\rho\otimes\omega)\M{C}_{ij}^\dagger \\
&\hspace{3mm}=\frac{1}{2}\left(\M{A}_i\M{B}_j\otimes \ket{0}\!\bra{0}+\M{B}_j\M{A}_i\otimes \ket{1}\!\bra{1}\right) \\
&\hspace{7mm}[\rho\otimes (\ket{0}\!\bra{0}+\ket{0}\!\bra{1}+\ket{1}\!\bra{0}+\ket{1}\!\bra{1})]\\
&\hspace{7mm}\left(\M{B}_j^\dagger\M{A}_i^\dagger\otimes \ket{0}\!\bra{0}+\M{A}_i^\dagger\M{B}_j^\dagger\otimes \ket{1}\!\bra{1}\right) \\
&\hspace{3mm}=\frac{1}{2}\Big(\M{A}_i\M{B}_j\rho\M{B}_j^\dagger\M{A}_i^\dagger\otimes \ket{0}\!\bra{0}+\M{A}_i\M{B}_j\rho\M{A}_i^\dagger\M{B}_j^\dagger\otimes \\
&\hspace{7mm}\ket{0}\!\bra{1}+\M{B}_j\M{A}_i\rho\M{B}_j^\dagger\M{A}_i^\dagger\otimes \ket{1}\!\bra{0}\\
&\hspace{7mm}+\M{B}_j \M{A}_i\rho\M{A}_i^\dagger\M{B}_j^\dagger\otimes \ket{1}\!\bra{1}\Big).
\end{aligned}
\end{equation}
Therefore, if we do not post-select on the control qubit, the partial state on the data register can be obtained by taking the partial trace as
\begin{equation}
\begin{aligned}
&\sum_{i,j}{\rm tr}_\omega\{\M{C}_{ij}(\rho\otimes\omega)\M{C}_{ij}^\dagger\}\\
&\hspace{3mm}=\frac{1}{2}\sum_{i,j}\M{A}_i\M{B}_j\rho\M{B}_j^\dagger\M{A}^\dagger
+\M{B}_j\M{A}_i\rho\M{A}_i^\dagger\M{B}_j^\dagger\\
&\hspace{3mm}=\frac{1}{4}\sum_{i,j}\{\M{A}_i,\M{B}_j\}\rho\{\M{A}_i,\M{B}_j\}^\dagger+[\M{A}_i,\M{B}_j]\rho[\M{A}_i,\M{B}_j]^\dagger.
\end{aligned}
\end{equation}
But if we discard the computational result once we measure a $\ket{-}$ on the control qubit, note that $\M{C}_{ij}(\rho\otimes\omega)\M{C}_{ij}^\dagger$ may be further expressed as

\begin{equation}
\begin{aligned}
&\hspace{-3mm}\M{C}_{ij}(\rho\otimes\omega)\M{C}_{ij}^\dagger \\
&=\frac{1}{4}\Big[\M{A}_i\M{B}_j\rho\M{B}_j^\dagger\M{A}_i^\dagger\!\otimes\!(
\rho_{++}+\rho_{+-}+\rho_{-+}+\rho_{--})\\
&\hspace{5mm}+\!\M{A}_i\M{B}_j\rho\M{A}_i^\dagger\M{B}_j^\dagger\!\otimes\!(
\rho_{++}-\rho_{+-}+\rho_{-+}-\rho_{--})\\
&\hspace{5mm}+\!\M{B}_j\M{A}_i\rho\M{B}_j^\dagger\M{A}_i^\dagger\!\otimes\!(
\rho_{++}+\rho_{+-}-\rho_{-+}-\rho_{--})\\
&\hspace{5mm}+\!\M{B}_j\M{A}_i\rho\M{A}_i^\dagger\M{B}_j^\dagger\!\otimes\!(
\rho_{++}-\rho_{+-}-\rho_{-+}+\rho_{--})\Big]\\
&=\frac{1}{4}\Big[\{\M{A}_i,\M{B}_j\}\rho\{\M{A}_i,\M{B}_j\}^\dagger\otimes \rho_{++} \\
&\hspace{7mm}+\{\M{A}_i,\M{B}_j\}\rho[\M{A}_i,\M{B}_j]^\dagger\otimes \rho_{+-} \\
&\hspace{7mm}+[\M{A}_i,\M{B}_j]\rho\{\M{A}_i,\M{B}_j\}^\dagger\otimes \rho_{-+} \\
&\hspace{7mm}+[\M{A}_i,\M{B}_j]\rho[\M{A}_i,\M{B}_j]^\dagger\otimes \rho_{--}\Big],
\end{aligned}
\end{equation}
where $\rho_{++}=\ket{+}\!\bra{+}$, $\rho_{+-}=\ket{+}\!\bra{-}$, $\rho_{-+}=\ket{-}\!\bra{+}$, $\rho_{--}=\ket{-}\!\bra{-}$. Hence the data register will be in the following state
\begin{equation}
\frac{1}{Z}\sum_{i,j}\frac{\{\M{A}_i,\M{B}_j\}}{2}\rho\frac{\{\M{A}_i,\M{B}_j\}^\dagger}{2},
\end{equation}
where $Z$ is a normalization factor given by
$$
Z = \frac{{\rm tr}\{\sum_{i,j}\{\M{A}_i,\M{B}_j\}\rho\{\M{A}_i,\M{B}_j\}^\dagger+[\M{A}_i,\M{B}_j]\rho[\M{A}_i,\M{B}_j]^\dagger\}}{{\rm tr}\{\sum_{i,j}\{\M{A}_i,\M{B}_j\}\rho\{\M{A}_i,\M{B}_j\}^\dagger\}}.
$$
This completes the proof.
\end{IEEEproof}

\bibliographystyle{ieeetran}
\bibliography{IEEEabrv,st_stab}

\begin{thebibliography}{10}
\providecommand{\url}[1]{#1}
\csname url@samestyle\endcsname
\providecommand{\newblock}{\relax}
\providecommand{\bibinfo}[2]{#2}
\providecommand{\BIBentrySTDinterwordspacing}{\spaceskip=0pt\relax}
\providecommand{\BIBentryALTinterwordstretchfactor}{4}
\providecommand{\BIBentryALTinterwordspacing}{\spaceskip=\fontdimen2\font plus
\BIBentryALTinterwordstretchfactor\fontdimen3\font minus
  \fontdimen4\font\relax}
\providecommand{\BIBforeignlanguage}[2]{{%
\expandafter\ifx\csname l@#1\endcsname\relax
\typeout{** WARNING: IEEEtran.bst: No hyphenation pattern has been}%
\typeout{** loaded for the language `#1'. Using the pattern for}%
\typeout{** the default language instead.}%
\else
\language=\csname l@#1\endcsname
\fi
#2}}
\providecommand{\BIBdecl}{\relax}
\BIBdecl

\bibitem{sycamore}
F.~Arute, K.~Arya, R.~Babbush, D.~Bacon, J.~C. Bardin, R.~Barends, R.~Biswas,
  S.~Boixo, F.~G. Brandao, D.~A. Buell \emph{et~al.}, ``Quantum supremacy using
  a programmable superconducting processor,'' \emph{Nature}, vol. 574, no.
  7779, pp. 505--510, 2019.

\bibitem{zuchongzhi}
\BIBentryALTinterwordspacing
Y.~Wu, W.-S. Bao, S.~Cao, F.~Chen, M.-C. Chen, X.~Chen, T.-H. Chung, H.~Deng,
  Y.~Du, D.~Fan \emph{et~al.}, ``Strong quantum computational advantage using a
  superconducting quantum processor,'' \emph{arXiv preprint}, 2021. [Online].
  Available: \url{https://arxiv.org/abs/2106.14734}
\BIBentrySTDinterwordspacing

\bibitem{fault_tolerance}
D.~Gottesman, ``Theory of fault-tolerant quantum computation,'' \emph{Phys.
  Rev. A}, vol.~57, no.~1, p. 127, 1998.

\bibitem{qecc}
A.~R. {Calderbank}, E.~M. {Rains}, P.~M. {Shor}, and N.~J.~A. {Sloane},
  ``Quantum error correction via codes over {GF$(4)$},'' \emph{IEEE Trans. Inf.
  Theory}, vol.~44, no.~4, pp. 1369--1387, Jul. 1998.

\bibitem{transversal}
E.~Knill, R.~Laflamme, and W.~H. Zurek, ``Resilient quantum computation: error
  models and thresholds,'' \emph{Proc. Roy. Soc. London A, Math. Phys. Eng.
  Sci.}, vol. 454, no. 1969, p. 365–384, Jan. 1998.

\bibitem{vqe}
P.~J. Love, J.~L. O'Brien, A.~Aspuru-Guzik, A.~Peruzzo, M.-h. Yung, X.-Q. Zhou,
  P.~Shadbolt, and J.~McClean, ``{A variational eigenvalue solver on a photonic
  quantum processor},'' \emph{Nature Commun.}, vol.~5, no.~1, pp. 1--7, Jul.
  2014.

\bibitem{vqe_theory}
J.~R. McClean, J.~Romero, R.~Babbush, and A.~Aspuru-Guzik, ``{The theory of
  variational hybrid quantum-classical algorithms},'' \emph{New Journal of
  Physics}, vol.~18, no.~2, pp. 1--22, Feb. 2016.

\bibitem{vqe2}
N.~Moll, P.~Barkoutsos, L.~S. Bishop \emph{et~al.}, ``{Quantum optimization
  using variational algorithms on near-term quantum devices},'' \emph{Quantum
  Science and Technology}, vol.~3, no.~3, pp. 1--17, Jun. 2018.

\bibitem{vqe_spectra}
T.~Jones, S.~Endo, S.~McArdle, X.~Yuan, and S.~C. Benjamin, ``Variational
  quantum algorithms for discovering {H}amiltonian spectra,'' \emph{Phys. Rev.
  A}, vol.~99, no.~6, Jun. 2019.

\bibitem{vqlinear}
X.~Xu, J.~Sun, S.~Endo, Y.~Li, S.~C. Benjamin, and X.~Yuan, ``Variational
  algorithms for linear algebra,'' \emph{Science Bulletin, \textit{early
  access}}, Jun. 2021.

\bibitem{sgd_vqa}
R.~Sweke, F.~Wilde, J.~J. Meyer, M.~Schuld, P.~K. F{\"a}hrmann,
  B.~Meynard-Piganeau, and J.~Eisert, ``Stochastic gradient descent for hybrid
  quantum-classical optimization,'' \emph{Quantum}, vol.~4, p. 314, 2020.

\bibitem{qaoa}
\BIBentryALTinterwordspacing
E.~Farhi, J.~Goldstone, and S.~Gutmann, ``A quantum approximate optimization
  algorithm,'' \emph{arXiv preprint}, 2014. [Online]. Available:
  \url{https://arxiv.org/abs/arXiv:1411.4028}
\BIBentrySTDinterwordspacing

\bibitem{performance_qaoa}
\BIBentryALTinterwordspacing
G.~E. Crooks, ``Performance of the quantum approximate optimization algorithm
  on the maximum cut problem,'' \emph{arXiv preprint}, 2018. [Online].
  Available: \url{https://arxiv.org/abs/1811.08419}
\BIBentrySTDinterwordspacing

\bibitem{scalable_simulation}
P.~J. O’Malley, R.~Babbush, I.~D. Kivlichan, J.~Romero, J.~R. McClean,
  R.~Barends, J.~Kelly, P.~Roushan, A.~Tranter, N.~Ding \emph{et~al.},
  ``Scalable quantum simulation of molecular energies,'' \emph{Phys. Rev. X},
  vol.~6, no.~3, 2016.

\bibitem{subspace_expansion1}
J.~I. Colless, V.~V. Ramasesh, D.~Dahlen, M.~S. Blok, M.~E. Kimchi-Schwartz,
  J.~R. McClean, J.~Carter, W.~A. de~Jong, and I.~Siddiqi, ``Computation of
  molecular spectra on a quantum processor with an error-resilient algorithm,''
  \emph{Phys. Rev. X}, vol.~8, Feb. 2018.

\bibitem{qem}
K.~Temme, S.~Bravyi, and J.~M. Gambetta, ``Error mitigation for short-depth
  quantum circuits,'' \emph{Phys. Rev. Lett.}, vol. 119, no.~18, pp. 1--5, Nov.
  2017.

\bibitem{qtc1}
D.~{Poulin}, J.~{Tillich}, and H.~{Ollivier}, ``Quantum serial turbo codes,''
  \emph{IEEE Trans. Inf. Theory}, vol.~55, no.~6, pp. 2776--2798, Jun. 2009.

\bibitem{qtc2}
Z.~{Babar}, S.~X. {Ng}, and L.~{Hanzo}, ``{EXIT}-chart-aided near-capacity
  quantum turbo code design,'' \emph{IEEE Trans. Veh. Technol.}, vol.~64,
  no.~3, pp. 866--875, Mar. 2015.

\bibitem{qtecc}
D.~{Chandra}, Z.~{Babar}, H.~V. {Nguyen}, D.~{Alanis}, P.~{Botsinis}, S.~X.
  {Ng}, and L.~{Hanzo}, ``Quantum topological error correction codes: {T}he
  classical-to-quantum isomorphism perspective,'' \emph{IEEE Access}, vol.~6,
  pp. 13\,729--13\,757, 2018.

\bibitem{qecc_survey}
Z.~{Babar}, D.~{Chandra}, H.~V. {Nguyen}, P.~{Botsinis}, D.~{Alanis}, S.~X.
  {Ng}, and L.~{Hanzo}, ``Duality of quantum and classical error correction
  codes: {D}esign principles and examples,'' \emph{IEEE Commun. Surv. Tuts.},
  vol.~21, no.~1, pp. 970--1010, 1st quart. 2019.

\bibitem{practical_qem}
S.~Endo, S.~C. Benjamin, and Y.~Li, ``Practical quantum error mitigation for
  near-future applications,'' \emph{Phys. Rev. X}, vol.~8, no.~3, pp. 1--21,
  Jul. 2018.

\bibitem{zne1}
T.~Giurgica-Tiron, Y.~Hindy, R.~LaRose, A.~Mari, and W.~J. Zeng, ``Digital zero
  noise extrapolation for quantum error mitigation,'' in \emph{Proc. IEEE Int.
  Conf. Quantum Comput. Engineering}, Virtual conference, Oct. 2020, pp.
  306--316.

\bibitem{zne2}
A.~He, B.~Nachman, W.~A. de~Jong, and C.~W. Bauer, ``Zero-noise extrapolation
  for quantum-gate error mitigation with identity insertions,'' \emph{Phys.
  Rev. A}, vol. 102, no.~1, p. 012426, Jan. 2020.

\bibitem{zne3}
\BIBentryALTinterwordspacing
Z.~Zhao and K.~C. Tan, ``Error mitigation in quantum metrology via zero noise
  extrapolation,'' \emph{arXiv preprint}, 2021. [Online]. Available:
  \url{https://arxiv.org/abs/2101.03766}
\BIBentrySTDinterwordspacing

\bibitem{qem_exp}
C.~Song, J.~Cui, H.~Wang, J.~Hao, H.~Feng, and Y.~Li, ``Quantum computation
  with universal error mitigation on a superconducting quantum processor,''
  \emph{Science Advances}, vol.~5, no.~9, 2019.

\bibitem{sof_analysis}
Y.~Xiong, D.~Chandra, S.~X. Ng, and L.~Hanzo, ``Sampling overhead analysis of
  quantum error mitigation: {U}ncoded vs. coded systems,'' \emph{IEEE Access},
  vol.~8, pp. 228\,967--228\,991, Dec. 2020.

\bibitem{ryuji_cost_qem}
\BIBentryALTinterwordspacing
R.~Takagi, ``Optimal resource cost for error mitigation,'' \emph{Phys. Rev.
  Research}, vol.~3, p. 033178, Aug. 2021. [Online]. Available:
  \url{https://link.aps.org/doi/10.1103/PhysRevResearch.3.033178}
\BIBentrySTDinterwordspacing

\bibitem{clifford_regression}
\BIBentryALTinterwordspacing
P.~Czarnik, A.~Arrasmith, P.~J. Coles, and L.~Cincio, ``Error mitigation with
  clifford quantum-circuit data,'' \emph{arXiv preprint}, 2021. [Online].
  Available: \url{https://arxiv.org/abs/2005.10189}
\BIBentrySTDinterwordspacing

\bibitem{sv}
X.~Bonet-Monroig, R.~Sagastizabal, M.~Singh, and T.~E. O'Brien, ``{Low-cost
  error mitigation by symmetry verification},'' \emph{Phys. Rev. A}, vol.~98,
  no.~6, pp. 1--10, Dec. 2018.

\bibitem{sv2}
J.~R. McClean, Z.~Jiang, N.~C. Rubin, R.~Babbush, and H.~Neven, ``Decoding
  quantum errors with subspace expansions,'' \emph{Nat. Commun.}, vol.~11,
  no.~1, pp. 1--9, 2020.

\bibitem{sv3}
\BIBentryALTinterwordspacing
Z.~Cai, ``Quantum error mitigation using symmetry expansion,'' \emph{arXiv
  preprint}, 2021. [Online]. Available: \url{https://arxiv.org/abs/2101.03151}
\BIBentrySTDinterwordspacing

\bibitem{sv4}
\BIBentryALTinterwordspacing
------, ``Resource-efficient purification-based quantum error mitigation,''
  \emph{arXiv preprint}, 2021. [Online]. Available:
  \url{https://arxiv.org/abs/2107.07279}
\BIBentrySTDinterwordspacing

\bibitem{balint_vd}
\BIBentryALTinterwordspacing
B.~Koczor, ``Exponential error suppression for near-term quantum devices,''
  \emph{arXiv preprint}, 2021. [Online]. Available:
  \url{https://arxiv.org/abs/2011.05942}
\BIBentrySTDinterwordspacing

\bibitem{vd2}
\BIBentryALTinterwordspacing
W.~J. Huggins, S.~McArdle, T.~E. O'Brien, J.~Lee, N.~C. Rubin, S.~Boixo, K.~B.
  Whaley, R.~Babbush, and J.~R. McClean, ``Virtual distillation for quantum
  error mitigation,'' \emph{arXiv preprint}, 2021. [Online]. Available:
  \url{https://arxiv.org/abs/2011.07064}
\BIBentrySTDinterwordspacing

\bibitem{qs_early}
Y.~Aharonov, J.~Anandan, S.~Popescu, and L.~Vaidman, ``Superpositions of time
  evolutions of a quantum system and a quantum time-translation machine,''
  \emph{Phys. Rev. Lett.}, vol.~64, pp. 2965--2968, Jun. 1990.

\bibitem{qs_comp}
G.~Chiribella, G.~M. D'Ariano, P.~Perinotti, and B.~Valiron, ``Quantum
  computations without definite causal structure,'' \emph{Phys. Rev. A},
  vol.~88, p. 022318, Aug. 2013.

\bibitem{qs_prl}
K.~Goswami, C.~Giarmatzi, M.~Kewming, F.~Costa, C.~Branciard, J.~Romero, and
  A.~G. White, ``Indefinite causal order in a quantum switch,'' \emph{Physical
  Rev. Lett.}, vol. 121, no.~9, p. 090503, Aug. 2018.

\bibitem{qs_ieee}
M.~Caleffi and A.~S. Cacciapuoti, ``Quantum switch for the quantum internet:
  {N}oiseless communications through noisy channels,'' \emph{IEEE J. Sel. Areas
  Commun.}, vol.~38, no.~3, pp. 575--588, 3rd Quart. 2020.

\bibitem{qs_comm1}
\BIBentryALTinterwordspacing
S.~Salek, D.~Ebler, and G.~Chiribella, ``Quantum communication in a
  superposition of causal orders,'' \emph{arXiv preprint}, 2018. [Online].
  Available: \url{https://arxiv.org/abs/1809.06655}
\BIBentrySTDinterwordspacing

\bibitem{qs_comm2}
G.~Chiribella and H.~Kristj{\'a}nsson, ``Quantum {S}hannon theory with
  superpositions of trajectories,'' \emph{Proc. Roy. Soc. A}, vol. 475, no.
  2225, p. 20180903, 2019.

\bibitem{qs_comm3}
P.~A. Gu{\'e}rin, G.~Rubino, and {\v{C}}.~Brukner, ``Communication through
  quantum-controlled noise,'' \emph{Phys. Rev. A}, vol.~99, no.~6, p. 062317,
  Jun. 2019.

\bibitem{qs_comm4}
G.~Chiribella, M.~Banik, S.~S. Bhattacharya, T.~Guha, M.~Alimuddin, A.~Roy,
  S.~Saha, S.~Agrawal, and G.~Kar, ``Indefinite causal order enables perfect
  quantum communication with zero capacity channels,'' \emph{New Journal of
  Physics}, vol.~23, no.~3, p. 033039, Mar. 2021.

\bibitem{spectral_analysis}
P.~Stoica, R.~L. Moses \emph{et~al.}, \emph{Spectral analysis of
  signals}.\hskip 1em plus 0.5em minus 0.4em\relax Pearson Prentice Hall Upper
  Saddle River, NJ, 2005, vol. 452.

\bibitem{perm_filter}
Y.~Xiong, S.~X. Ng, and L.~Hanzo, ``Quantum error mitigation relying on
  permutation filtering,'' \emph{IEEE Trans. Commun.}, vol.~70, no.~3, pp.
  1927--1942, Mar. 2022.

\bibitem{dm}
L.~Wang, L.~Li, C.~Xu, D.~Liang, S.~X. Ng, and L.~Hanzo, ``Multiple-symbol
  joint signal processing for differentially encoded single- and multi-carrier
  communications: {P}rinciples, designs and applications,'' \emph{IEEE Commun.
  Surv. Tuts.}, vol.~16, no.~2, pp. 689--712, 2nd Quart. 2014.

\bibitem{dstm1}
B.~Hughes, ``Differential space-time modulation,'' \emph{IEEE Trans. Inf.
  Theory}, vol.~46, no.~7, pp. 2567--2578, Jul. 2000.

\bibitem{dstm2}
B.~Hochwald and W.~Sweldens, ``Differential unitary space-time modulation,''
  \emph{IEEE Trans. Commun.}, vol.~48, no.~12, pp. 2041--2052, Dec. 2000.

\bibitem{ncbook}
M.~A. Nielsen and I.~L. Chuang, \emph{Quantum Computation and Quantum
  Information}, 2nd~ed.\hskip 1em plus 0.5em minus 0.4em\relax New York, NY,
  USA: Cambridge University Press, 2011.

\bibitem{shor}
P.~W. {Shor}, ``Algorithms for quantum computation: {D}iscrete logarithms and
  factoring,'' in \emph{Proc. 35th Annual Symp. Foundations of Computer
  Science}, Santa Fe, New Mexico, USA, Nov. 1994, pp. 124--134.

\bibitem{hhl}
A.~W. Harrow, A.~Hassidim, and S.~Lloyd, ``Quantum algorithm for linear systems
  of equations,'' \emph{Phys. Rev. Lett.}, vol. 103, no.~15, Oct. 2009.

\bibitem{ibmq}
\BIBentryALTinterwordspacing
``{IBM} quantum compute resources,'' {A}ccessed: Aug. 10, 2022. [Online].
  Available: \url{https://quantum-computing.ibm.com/services/resources}
\BIBentrySTDinterwordspacing

\bibitem{questlink}
\BIBentryALTinterwordspacing
T.~Jones and S.~C. Benjamin, ``{QuEST}link -- {M}athematica embiggened by a
  hardware-optimised quantum emulator,'' \emph{arXiv preprint}, 2019. [Online].
  Available: \url{https://arxiv.org/abs/1912.07904}
\BIBentrySTDinterwordspacing

\end{thebibliography}
\end{document}